\let\TeXyear\year
\documentclass{ieeeaccess}

\let\year\TeXyear
\usepackage[
  backend=biber,
  maxcitenames=3,
  minbibnames=1,
  mincitenames=1,
  maxbibnames=3,
  uniquelist=false,
  url=false,
  doi=true,
  eprint=false,
  isbn=false,
]{biblatex}
\usepackage{amsmath,amssymb,amsfonts}
\usepackage{algorithmic}
\usepackage{algorithm}
\usepackage{graphicx}
\usepackage{textcomp}
\usepackage{xcolor}
\usepackage{subfig}
\usepackage{braket}
\usepackage{comment}
\usepackage{csquotes}
\usepackage{xspace}
\usepackage{tikz}
\usepackage{pgfplots}
\usepackage{tabularray}
\UseTblrLibrary{booktabs}
\pgfplotsset{compat=1.18}
\NewSpotColorSpace{PANTONE}
  \AddSpotColor{PANTONE} {PANTONE3015C} {PANTONE\SpotSpace 3015\SpotSpace C} {1 0.3 0 0.2}
  \SetPageColorSpace{PANTONE}%
\usepackage[T1]{fontenc}
\usepgfplotslibrary{groupplots}
\usepgfplotslibrary{colormaps}

\usepackage[colorlinks=true,urlcolor=teal,
            linkcolor=teal,citecolor=teal]{hyperref}

\newcommand{\email}[1]{\href{mailto:#1}{#1}}

\newcommand{\ie}{\emph{i.e.}\xspace}
\newcommand{\eg}{\emph{e.g.}\xspace}
\newcommand{\cf}{cf.\xspace}
\newcommand{\etal}{\emph{et al.}\xspace}
\newcommand{\reprolink}{https://github.com/lfd/QAOA_non_patterns_2025}
\newcommand{\repro}{\href{\reprolink}{reproduction package}\xspace}
\newcommand*{\I}{\imath} 

\newcommand{\COBYLA}{COBYLA}

\newcommand{\LRm}{\ensuremath{\text{LR}_{-\beta}}\xspace}
\newcommand{\LRp}{\ensuremath{\text{LR}_{+\beta}}\xspace}
\newcommand{\LR}{\ensuremath{\text{LR}}\xspace}

\definecolor{lfd1}{HTML}{FFFFFF} 
\definecolor{lfd2}{HTML}{E69F00}
\definecolor{lfd3}{HTML}{999999}
\definecolor{lfd4}{HTML}{009371}

\definecolor{lfd5}{HTML}{BEAED4}
\definecolor{lfd6}{HTML}{ED665A}
\definecolor{lfd7}{HTML}{1F78B4}

\begin{document}
\doi{XXX}

\title{Going off Pattern? QAOA Parameter Heuristics and Potentials of Parsimony}
\author{\uppercase{Vincent Eichenseher}\authorrefmark{1},
\uppercase{Maja Franz}\authorrefmark{1},
\uppercase{Christian Wolff}\authorrefmark{2},
\uppercase{Wolfgang Mauerer}\authorrefmark{1,3}}

\address[1]{Technical University of Applied Sciences Regensburg, Regensburg, Germany}
\address[2]{University of Regensburg, Regensburg, Germany}
\address[3]{Siemens AG, Foundational Technologies, Munich, Germany}

\newcommand{\VE}{VE\xspace}
\newcommand{\MF}{MF\xspace}
\newcommand{\CW}{CW\xspace}
\newcommand{\WM}{WM\xspace}
\newcommand{\programme}{German Federal Ministry of Research, Technology and Space (BMFTR), funding program \enquote{Quantum Technologies~--~from
      Basic Research to Market}}
\newcommand{\grantoth}{\#13NI6092}
\newcommand{\hta}{High-Tech Agenda Bavaria}

\tfootnote{We acknowledge support from \programme, grant \grantoth{} (\VE, \MF and \WM). \WM acknowledges support by the \hta.}

\corresp{Corresponding authors: Vincent Eichenseher (\email{vincent.eichenseher@othr.de}), Maja Franz (\email{maja.franz@othr.de})}

\begin{abstract}
Structured variational quantum algorithms such as the Quantum Approximate Optimisation Algorithm (QAOA) have emerged as leading candidates for exploiting advantages of near-term quantum hardware. They interlace classical computation, in particular optimisation of variational parameters, with quantum-specific routines, and combine problem-specific advantages~--~sometimes even provable~--~with adaptability to the constraints of noisy, intermediate-scale quantum (NISQ) devices. While circuit depth can be parametrically increased and is known to improve performance in an ideal (noiseless) setting, on realistic hardware greater depth exacerbates noise: The overall quality of results depends critically on both, variational parameters and circuit depth. Although identifying optimal parameters is NP-hard, prior work has suggested that they may exhibit regular, predictable patterns for increasingly deep circuits and depending on the studied class of problems.

In this work, we systematically investigate the role of classical parameters in QAOA performance through extensive numerical simulations and suggest a simple, yet effective heuristic scheme to find good parameters for low-depth circuits. Our results demonstrate that: (i) high-quality parameters often deviate substantially from expected patterns; (ii) QAOA performance becomes progressively less sensitive to specific parameter choices as depth increases; and (iii) iterative component-wise fixing performs on par with, and at shallow depth may even outperform, several established parameter-selection strategies. We identify conditions under which structured parameter patterns emerge, and when deviations from the patterns warrant further consideration.
These insights for low-depth circuits may inform more robust pathways to harnessing QAOA in realistic quantum computing scenarios. 
\end{abstract}


\maketitle
\renewcommand{\emph}[1]{\textit{#1}}
\section{Introduction}\label{sec:introduction}
Despite progress towards error-corrected quantum
computers~\cite{google_willow24}, noise and imperfections remain dominant in NISQ devices~\cite{Bharti:2022,gardas_defects_2018}. Moreover, algorithms with proven quantum advantage~\cite{kim_evidence_2023} are impractical at current qubit scales.
Over the past decade, research has centred on hybrid quantum–classical algorithms adapted to restricted hardware. These methods replace full coherent quantum evolution with iterative classical protocols that invoke quantum subroutines, thereby reducing qubit and gate requirements~\cite{peruzzo_variational_2014, safi23}. This mitigation limits noise accumulation on small-scale NISQ implementations. Such variational quantum algorithms (VQAs) \cite{cerezo_variational_2021} include the variational quantum eigensolver (VQE)~\cite{peruzzo_variational_2014} and the quantum approximate optimisation algorithm
(QAOA)~\cite{farhi_quantum_2014}.

QAOA admits variable circuit depths by repeating parametrised mixer and cost unitaries, that is, the number of layers \(p\). As the algorithm corresponds
to a Trotterisation of adiabatic quantum evolution~\cite{Bharti:2022},
the corresponding dynamics are recovered in the limit \(p \to \infty\) and yields the optimal solution (or a optimal solution, in the case that there are degeneracies) in the absence of noise.

Identifying optimal parameters for finite \(p\) is known to be NP-hard~\cite{bittel_training_2021}. Nonetheless, for small and modest depths, both empirical and theoretical studies indicate that good approximations are feasible for many problems~\cite{Bharti:2022, Carbonelli2024, schoenberger23, schmidbauer25, gogeiss24}. While classical optimisation remains the most common strategy \cite{blekos_review_2024,fernandez-pendas_study_2022,lotshaw_empirical_2021}, alternative approaches have been proposed that yield high-quality parameters at reasonable computational cost \cite{Tate:2023,Gavon:2024,Egger:2021,Vijendran:2024,Sud:2024,krueger25}.

Several studies suggest the existence of discernible patterns in optimal QAOA parameters~\cite{zhou_quantum_2020,lee_parameters_2021,montanez-barrera_towards_2024}. Typically, cost-unitary parameters are observed to increase smoothly with depth, while mixer-unitary parameters decrease. Empirical results further indicate that parameters at depth \(p\) differ only slightly from those at \(p+1\)~\cite{zhou_quantum_2020}. This behaviour aligns with the fact that larger \(p\) correspond to a more fine-grained Trotterisation of adiabatic time evolution. 
In this work, we re-examine the assumption that optimal parameters adhere to such patterns by systematically investigating the cost landscapes of optimisation-free parameter initialisation methods for the canonical NP-complete problems \emph{MaxCut}, \emph{VertexCover}, and \emph{Max3SAT}.
In particular, we examine linear ramp schedules~\cite{montanez-barrera_towards_2024} and compare their performance with parameters obtained through classical optimisation and a brute-force sequential method.
Our analysis includes not only parameters consistent with the established patterns, but more importantly seeks systematic deviations, thereby accounting for degenerate optima and relating varying degrees of pattern-conformity to the expected solution quality. 
This enables an assessment of the extent to which optimal parameters realistically follow proposed patterns. In contrast, prior approaches typically focus solely on parameters identified by optimisation routines or other selection methods, offering no insight into the performance of neighbouring points in the optimisation landscape dynamics~\cite{albash_adiabatic_2018}.
In summary, our main contributions are:

\begin{enumerate}
  \item We replicate and confirm~--~using a comprehensive open source \repro~--~results reported in the literature, and further analyse the influence of component-wise parameter adjustments on corresponding cost-landscapes. We employ large-scale numerical simulations for deep circuits beyond 20 layers.
  \item We examine the conformance of optimal QAOA parameters to patterns suggested in the literature, 
  and show that actually optimised parameters in practical settings often \emph{fail} to follow these.
  We provide guidance on when such patterns are likely to hold, as well as when and to what degree deviations may arise. We identify possible reasons, and uncover more accurate patterns.
  \item We report on extensive experiments for a \emph{sequential} method to iterative parameter fixing in QAOA and perform a comparative analysis with other well known methods, demonstrating that it yields, despite its structural simplicity and little required computational effort, comparatively good results akin to more complex schemes at low depths.
\end{enumerate}

The rest of this article is structured as follows. \autoref{sec:theory} outlines the theoretical background and terminology. Related work is reviewed in \autoref{sec:rel_work}, followed by the experimental setup in \autoref{sec:method}. Results are presented in \autoref{sec:results}, with attention to symmetries, parameter concentration, and fixed-parameter implications. \autoref{sec:discussion} considers the potential of optimisation-free approaches and the role of parameter patterns. We conclude in \autoref{sec:conclusion}.

\section{Foundations}\label{sec:theory}
QAOA~\cite{farhi_quantum_2014} effectively describes the Trotterisation of quantum adiabatic annealing~\cite{farhi_quantum_2000}, which seeks to determine the ground state of a problem Hamiltonian by continuously interpolating between an easy to construct initial Hamiltonian and the problem Hamiltonian, exploiting adiabatic dynamics to ensure that the system remains in the ground state.
In QAOA the continuous annealing time $T$ in adiabatic quantum computation is replaced with $p$ discrete time steps, which allows to implement the algorithm on gate-based hardware.
For finite time steps $p$ the QAOA presents an approximation of the ideal time evolution, which can be employed to also approximate solutions to unconstrained optimisation problems by measuring the quantum state obtained after applying \(p\) layers of alternating unitaries to the initial state 
\(\ket{s} = \ket{+}^{\otimes n}\).

Each layer \(i \in [1,p]\) applies a phase-separation operator
\begin{equation}\label{eq:phase-operator}
  H_C = \sum_{i,j} J_{ij}\sigma_z^{(i)} \sigma_z^{(j)} + \sum_{i} h_i \sigma_z^{(i)},
\end{equation}
where \(J_{ij}\) specifies interaction strengths and \(h_i\) denotes local energy offsets. Classical variables are encoded via \(\sigma_z^{(i)}\). This is followed by a mixer unitary \(U_B(\beta_i) = e^{-\I \beta_i H_B}\) that commutes with the phase-separation operator. 
When \(\ket{s}\) is a product of \(\sigma_x\) eigenstates, the transverse-field Hamiltonian \(H_B = \sum_{i=0}^n \sigma_x^{(i)}\) is typically chosen for this purpose.

The repeated application of these layers results in the parametrised quantum state
\begin{equation}\label{eq:qaoa_parameterized_state}
     \ket{\gamma,\beta} = U_{B}(\beta_{p})U_{C}(\gamma_{p})\cdots U_{B}(\beta_{1})U_{C}(\gamma_{1})\ket{s},
\end{equation}
that depends on $2p$ real parameters $\vec{\gamma} =(\gamma_{1},\gamma_{2},\cdots,\gamma_{p})$ and $\vec{\beta} = (\beta_{1},\beta_{2},\cdots,\beta_{p})$.
Parameter optimisation seeks parameters $\vec{\gamma}, \vec{\beta}$ that minimise (or maximise) the expectation value 
\begin{equation}
\label{eq:expectation_value}
     F_{p}(\vec{\gamma}, \vec{\beta}) = \bra{\vec{\gamma}, \vec{\beta}}H_C\ket{\vec{\gamma}, \vec{\beta}}
\end{equation}
that can be sampled from the quantum circuit implementing \autoref{eq:expectation_value}.
Optimisation is usually performed by classical iterative methods that produce a sequence of gradually improving choices for $(\vec{\gamma}, \vec{\beta})$. 

For most problems, \(\vec{\gamma}\) is \(2\pi\)-periodic and using the transverse-field Hamiltonian as a mixer, \(\vec{\beta}\) is \(\pi\)-periodic. We refer to these periodicities as \enquote{common symmetries}. The expectation value is also invariant under time reversal,
\(F_p(\vec{\gamma},\vec{\beta}) = F_p(-\vec{\gamma},-\vec{\beta})\), and problem-specific symmetries may further restrict the parameter space. For MaxCut on regular graphs, \(\beta \in [-\pi/4, \pi/4]\) and \(\gamma \in [-\pi/2, \pi/2]\), which we refer to as \enquote{MaxCut symmetries}, whereas for non-regular graphs~--~or other optimisation problems~--~only the common symmetry applies, so \(\gamma \in [-\pi, \pi]\).
Note that different periods may arise for the cost operator parameter in some problems (dependant on the distributions of objective values) and different mixer-periods may apply when using different mixers.

\label{sec:rcc}
The expectation value in \autoref{eq:expectation_value} can be decomposed into a sum of terms that involve only qubits $i$ and $j$:
\begin{equation}
\label{eq:decomposed_expectation_value}
  F_p(\vec{\gamma}, \vec{\beta}) = \sum_{i,j} J_{ij} \bra{\vec{\gamma}, \vec{\beta}}\sigma_z^{(i)}\sigma_z^{(j)}\ket{\vec{\gamma}, \vec{\beta}}.
\end{equation}

The collection of terms involving qubits \(i\) or \(j\) constitutes the \emph{reverse causal cone} of their correlation function. For each depth \(p\), this cone contains only finitely many terms~\cite{Streif:2020}, and its size depends solely on \(p\), independent of the instance size \(n\).

\section{Related Work}\label{sec:rel_work}
QAOA was introduced by Farhi~\etal in 2014~\cite{farhi_quantum_2014}, and has since inspired numerous variants and extensions~\cite{Hadfield:2019,bravyi_obstacles_2020,wurtz_counterdiabaticity_2022,wurtz_classically_2021,Tate:2023}.
The idea of Trotterising adiabatic quantum time evolution predates QAOA (see \eg, \cite{wu_polynomial_2002}), but initially targeted Hamiltonian simulation. The general idea of using adiabatic quantum computing for solving combinatorial optimisation problems~\cite{farhi_quantum_2000}, was then first realised as the QAOA in Ref.~\cite{farhi_quantum_2014}.
As simulating the output of QAOA is classically intractable even for depth \(p=1\)~\cite{farhi_quantum_2019} given some widely accepted complexity-theoretic assumptions, it is frequently seen as indicator of quantum advantage. While the single-layer case is well understood~\cite{krueger25,farhi_quantum_2014,bravyi_obstacles_2020,galda_transferability_2021}, the behaviour of deeper circuits remains less clear, notwithstanding a steadily expanding body of results~\cite{zhou_quantum_2020,wurtz_maxcut_2021,wurtz_fixed_2021,lee_parameters_2021,lee_depth-progressive_2023}.

In particular, this includes the identification of patterns across problem instances~\cite{Streif:2020,montanezbarrera:2024}, for instance through machine learning~\cite{Jain:2022},
analytical considerations~\cite{krueger25}, or by other  approaches~\cite{zhou_quantum_2020,lee_parameters_2021}. Such patterns aim to provide less computationally demanding alternatives than orthodox optimisation to determine good parameters \(\vec{\beta}, \vec{\gamma}\). In general, standard approaches are costly and require multiple quantum circuit evaluations, although the  exact effort depends on many details~\cite{periyasamy:24:qce24,fernandez-pendas_study_2022,tibaldi_bayesian_2023}.

Zhou~\etal~\cite{zhou_quantum_2020} observed that optimal parameters exhibit smooth trends, with \(\gamma_i\) increasing and \(\beta_i\) decreasing as functions of the layer index \(i \in [1,p]\). Exploiting these patterns, they proposed to let parameters obtained at low depth serve as initial values for deeper circuits, either via interpolation or through low-depth amplitudes in a frequency-domain approximation. Both approaches were shown to outperform random initialisation of QAOA.
In a recent work by Apte~\etal~\cite{apte_iterative_2025}, the authors generalise the parameter selection methods of Zhou~\etal~\cite{zhou_quantum_2020}, noting that smooth QAOA schedules can be decomposed into a basis of orthonormal functions
The approach used by Zhou~\etal presents a special case of this approach using trigonometric functions.
In Ref.~\cite{apte_iterative_2025} the authors note that smooth schedules can be well approximated with a small number of basis coefficients and that optimising a smaller number of coefficients reduces optimisation complexity.
The proposed method iteratively increases both the circuit depth and number of coefficients used in the orthonormal basis expansion until the algorithm converges, and demonstrate that the method achieves better results in fewer function evaluations than the methods proposed by Zhou~\etal.

Other iterative methods for obtaining good higher depth parameters include the parameter fixing strategy proposed by Lee~\etal in ~\cite{lee_parameters_2021}, further refined in Ref.~\cite{lee_depth-progressive_2023}, which initialise the classical optimiser with the best parameters found at a previous depth.
In particular, they initialise a QAOA circuit with the best parameters from a lower depth, optimise the full parameter set again and fix the best parameters for the next depth. This process is repeated until a target depth is reached.
The authors have shown that this strategy was able to achieve high quality results for MaxCut.
Pelofske~\etal~\cite{pelofske_scaling_2024} investigated the iterative fixing of parameters by performing grid-searches over only the highest depth components of $\beta,\gamma$ in each iteration, which is similar to our sequential method. They noted that the expectation value of the landscapes evaluated by these grid-searches begin to converge at low depth ($p\leq4$), showing little improvement between $p=3$ and $p=4$. The mechanisms behind this was not further investigated in their work, as the authors chose to use a different approach for their main results on QAOA parameter transfer (\ie, JuliQAOA~\cite{golden_juliqaoa_2023}). The technique used by JuliQAOA for extrapolating parameters uses parameter fixing with basin-hopping~\cite{wales_global_1997} to explore nearby local minima, and was shown to outperform the empirical performance bounds of random local minima exploration and median angles approaches shown in Ref.~\cite{lotshaw_empirical_2021}.

Brandão~\etal~\cite{brandao_for_2018} demonstrated that parameters concentrate for related problem instances (\eg, MaxCut on three-regular graphs), with the variance of optimal parameters decreasing as instance size increases. This suggests that either sufficiently large instances or averages over many smaller instances are necessary to identify a meaningful concentration point. By formally considering infinite system size through restriction to the reverse causal cone, Streif and Leib~\cite{Streif:2020} further showed that parameter concentration does not depend on the global problem size \(n\), but rather on qubit correlations. Galda~\etal~\cite{galda_transferability_2021} investigated the transferability of \(p=1\) QAOA parameters based on the types of subgraphs in the problem graph, and observed good transferability of parameters when both subgraphs are either even-degree or odd-degree, and poor transferability for pairs of even- and odd-degree subgraphs, and demonstrate how to identify small instances for which optimal parameters can be transferred from to large instances with little loss in approximation quality. Aside from MaxCut, there is evidence for parameter concentration for several other problems, including the Sherrington-Kirkpatrick model at infinite size~\cite{farhi_quantum_2022}, Max-kXOR~\cite{chou_limitations_2022}, the low autocorrelation binary sequences problem~\cite{shaydulin_evidence_2024} and higher order Ising models~\cite{pelofske_quantum_2023,pelofske_short-depth_2024,pelofske_scaling_2024}.

Given the optimal total time $T^*$ used in an adiabatic evolution, QAOA parameters can be initialised as linear schedules controlled by the Trotter time step $\Delta t=T^*/p$ by setting $\gamma_i=\frac{i}{p}\Delta t$ and $\beta_i=(1-\frac{i}{p})\Delta t$.
The results of this initialisation strategy, known as Trotterised Quantum Annealing (TQA), have been shown to be qualitatively similar to the best results obtained from an exponentially scaling number of random initialisations \cite{sack_quantum_2021}.
Refs.~\cite{kremenetski_quantum_2021,montanezbarrera:2024,montanez-barrera_towards_2024} suggest that parameters initialised via linear ramp (LR) schedules, with \(\gamma\) increasing linearly and \(\beta\) decreasing linearly, generalise effectively across different problems and instances. Such schedules also provide robust initialisations for subsequent optimisation~\cite{montanezbarrera:2024}.  

\section{Method}\label{sec:method}
In the following we present our sequential parameter initialisation method. Furthermore, we outline the considered metrics and optimisation problems, as well as our experimental setup.

\subsection{Sequential Parameter Initialisation}
\label{sec:sequential}

\begin{figure*}[htbp]
  \centering
  \input{method_illustration}
 \caption{Sequential parameter initialisation method, introduced in Ref.~\cite{pelofske_quantum_2023}.}
  \label{fig:method}
\end{figure*} 

As a baseline for assessing alternative optimisation and initialisation strategies, we propose a \emph{sequential} parameter optimisation scheme: QAOA parameters \((\beta,\gamma)\) are sampled from a uniform \(32 \times 32\) grid over \([-\frac{\pi}{2},\frac{\pi}{2}] \times [-\pi,\pi]\), or, where symmetry permits, over \([-\frac{\pi}{4},\frac{\pi}{4}] \times [-\frac{\pi}{2},\frac{\pi}{2}]\) (see \autoref{subsec:symmetries}). The procedure begins at depth \(p=1\), with all \(32^2\) grid points evaluated.
At each depth \(p\), the parameters yielding the lowest energy expectation are fixed. The depth is then incremented to \(p+1\), and the same parameter grid is explored with previously fixed components held constant. This process, which is illustrated in \autoref{fig:method}, continues until the target depth \(p_{\text{target}}\) is reached.
In this way, only a restricted region of the optimisation landscape close to the chosen fixed parameters requires evaluation, leading to linear rather than exponential scaling of the number of QAOA evaluations with depth \(p\). Although conceptually simple, fixing the best available tuple \((\beta,\gamma)\) at each stage may be suboptimal, as the effect of additional layers on the energy landscape remains uncertain.

As mentioned in \autoref{sec:rel_work}, the strategy of fixing lower-level QAOA parameters while iteratively increasing circuit depth has precedent, however unlike approaches which use an optimiser and re-optimise all components of \((\beta,\gamma)\) each iteration (see e.g.~\cite{lee_parameters_2021,lee_depth-progressive_2023,golden_juliqaoa_2023}), our  approach is more similar to the iterative grid-searches with parameter fixing investigated in~\cite{pelofske_scaling_2024}, which leaves lower depth components fixed and only searches over the highest-depth components in each iteration.
Differences in our implementation include a finer grid (containing 1024 parameters in our approach, 200 in~\cite{pelofske_scaling_2024}) and the fact that we use a full statevector simulation without finite sampling in our experiments (while~\cite{pelofske_scaling_2024} use 10000 shots).
However, most notably, we give a frame of reference for how the energy landscapes of sequentially fixed parameters compare to those of other methods, which are detailed in \autoref{sec:exp-setup}.
Here, we perform additional grid evaluations, where we substitute the fixed parameters with parameters given by these other well-known parameter selection methods.
This way, we aim to more thoroughly study the convergence mechanism noted by the authors in~\cite{pelofske_scaling_2024}.
Furthermore, we perform extensive experiments for greater circuit depths and a wider range of combinatorial problems and problem instances, outlined in \autoref{sec:cops}, than those reported on in ~\cite{pelofske_scaling_2024}.

\subsection{Metrics}
We assess the performance of a given set of parameters \((\vec{\gamma},\vec{\beta})\) using the energy expectation value \(F_p(\vec{\gamma},\vec{\beta})\) (\cf \autoref{eq:expectation_value}).
To enable comparison across problem instances, we employ the residual energy
\begin{equation}
r = \frac{F(\beta,\gamma)-E_0}{E_\text{max}-E_0},
\end{equation}
as a metric, where \(E_0\) and \(E_\text{max}\) denote the ground- and maximum-excited-state energies of the problem Hamiltonian \(H_C\). For multiple instances, we compute the mean residual energy \(\bar{r}\) and its standard deviation \(\sigma_r\). The ground state energy \(E_0\) is obtained by exactly diagonalising the Hamiltonian using the Numpy's eigensolver~\cite{harris2020array}.
While broadly consistent with other quality measures, such as the energy approximation ratio, the residual energy is particularly suited to evaluating states that are good but not optimal~\cite{Streif:2020}.

\subsection{Combinatorial Problems}
\label{sec:cops}
The QAOA can be employed to solve quadratic unconstrained binary optimisation (QUBO) problems, to which any optimisation problem in NP can be reduced~\cite{lucas14}.
The objective here is to minimise a cost function given by
\begin{equation}
  C(\vec{x}) = \vec{x}^T Q \vec{x},
\end{equation}
for a QUBO $Q \in \mathbb{R}^{n \times n}$, and $n$ binary variables $\vec{x}^T = (x_1, x_2, \dots, x_n) \in \{0,1\}^n$.
By switching from a binary variable $x_i \in \{0,1\}$ to a spin variable $s_i = 1 - 2x_i \in \{-1,1\}$~\cite{lucas14}, a QUBO can equivalently be described by an Ising model $\hat{H}_C$ that can be implemented in the QAOA.

In this work we consider three specific seminal NP-complete problems: MaxCut, VertexCover and Max3SAT.

\subsubsection{MaxCut}
MaxCut seeks a partition of a graph $G = (V, E)$ with edges $E$ and vertices $V$ that maximises the number of crossing edges~\cite{karp_reducibility_1972}. In the context of QAOA, MaxCut is one of the best-studied optimisation problems, with some assumptions for improvements over classical heuristics~\cite{farhi_quantum_2014}. The QUBO cost function can be formulated as~\cite{glover_quantum_2022}:
\begin{equation}
C_\text{MC}(\vec{x}) = \sum_{(i,j)\in E} (- x_i - x_j + 2x_ix_j).
\end{equation}
Our study considers three-regular graphs, consistent with the QAOA literature, where \(d\)-regular graphs are standard \cite{zhou_quantum_2020,brandao_for_2018,fernandez-pendas_study_2022,wurtz_fixed_2021,wurtz_maxcut_2021,farhi_quantum_2014}. We evaluated 40 instances in total: ten each of orders 10, 12, 14, and 16. We chose these problem sizes in accordance with \cite{zhou_quantum_2020}, who noted the presence of patterns for three-regular graph instances of these sizes. Furthermore, many studies, which propose patterns in optimal QAOA parameters (discussed in \autoref{sec:rel_work}), investigate instances of similar size (see e.g. \cite{montanez-barrera_towards_2024,lee_parameters_2021,lee_depth-progressive_2023}).

\subsubsection{VertexCover}
VertexCover seeks a minimal set of nodes in a graph such that every edge is adjacent to at least one node in the set~\cite{karp_reducibility_1972}.
The constrained version of this problem can be formulated as \(\min \sum_{j\in V} x_i \;\text{subject to}\; x_i + x_j\geq1 \forall (i,j)\in E \) which we convert into unconstrained form by representing the constraints with penalty terms, resulting in the QUBO formulation~\cite{glover_quantum_2022}:
\begin{equation}
C_\text{VC}(\vec{x}) = \sum_{i\in V} x_i + P\left(\sum_{(i,j)\in E}(1-x_j-x_i+x_ix_j)\right),
\end{equation}
where \(P\) is a positive scalar penalty value, \(V\) is the set of vertices and \(E\) is again the set of edges \((i,j)\).
For our experiments, we employ the same three-regular graph instances as in the MaxCut study: ten each of orders 10, 12, 14, and 16. Reusing the structural properties of the graphs, yet changing the problem itself allows to investigate the influence of the specific objectives.

\subsubsection{Max3SAT}
Max3SAT involves assigning Boolean variables in a 
conjunctive normal form (CNF) formula (with three Boolean variables per clause, 3-CNF) so as to maximise the number of satisfied clauses~\cite{karp_reducibility_1972,mitchell_hard_1992}.
To transform a Max3Sat problem in 3-CNF, \(\land_{i=0}^m(a\lor b\lor c)\)with \(a,b,c\) chosen from binary variables \( x \in\{0,1\}^n \), we follow the approach of Chancellor \etal~\cite{chancellor_direct_2016}.
Here, the individual clauses are mapped to QUBOs, such that all satisfying assignments have the lowest energy.
Subsequently, these individual QUBOs are superimposed to a common basis, yielding a combined QUBO which represents our problem (see alse Refs.~\cite{zielinski_pattern_2023, schmidbauer25_sat} for a more detailed description of Chancellor \etal's method).
Since the QUBOs for the individual clauses contain cubic terms, we have to add auxiliary variables \( \vec{y} \in\{0,1\}^m \) to model these terms with quadratic interactions.
Specifically, cubic terms \(x_i x_j x_k\) are modelled by replacing a product of two variables \(x_i x_j\) with an auxiliary variable \(y_{ij}\).
Using this mapping, a formula with \(n\) variables and \(m\) clauses is represented by \(n+m\) qubits and can be formalised as
\begin{equation}
\min\left\{(\vec{x},\vec{y})^\mathrm{T} Q (\vec{x},\vec{y})\right\}.
\end{equation}
The clause-specific QUBOs superimposed to construct \(Q\) are listed in \autoref{tab:max3sat_clauses}, with the penalty term \(P(x_ix_j -2x_iy_{ij} - 2x_jy_{ij} + 3y_{ij})\), where \(P\) is a positive scalar penalty~\cite{boros_pseudo-boolean_2002,verma_efficient_2021}.

\begin{table}[]
    \centering
    \begin{tblr}{width=\linewidth,
            colspec={lX[l]},
            row{2-Z}={belowsep=0.15em}}
              \toprule
              \textbf{Type of Clause} & \textbf{Objective} \\
              \midrule
              $(x_i\lor x_j\lor x_k)$ & $-x_i-x_j-x_k+x_ix_j+x_ix_k+x_jx_k-x_ix_jx_k $\\
              $(x_i\lor x_j\lor \neg x_k)$ & $-1+x_k-x_ix_k-x_jx_k+x_ix_jx_k$ \\
              $(x_i\lor \neg x_j\lor \neg x_k)$ & $-1-x_jx_k-x_ix_jx_k$ \\
              $(\neg x_i\lor \neg x_j\lor \neg x_k)$ & $-1-x_ix_jx_k$ \\
              \bottomrule
    \end{tblr}
    \caption{Clause-specific QUBOs, from which the Max3SAT problem QUBO is constructed.}
    \label{tab:max3sat_clauses}
\end{table}
\vspace{0.5em}

Instances are characterised by the clause-to-variable ratio $\alpha=\lvert\text{Clauses}\rvert/\lvert\text{Variables}\rvert=m/n$. For small \(\alpha\), formulas are typically under-constrained with many solutions; for large \(\alpha\), they become over-constrained and unsatisfiable. Hard instances concentrate in \(\alpha \in (3.5, 4.9]\), where satisfiability occurs with a probability of about 50\%. The critical point shifts with formula size, converging near \(\alpha \approx 4.25\) for large systems~\cite{mitchell_hard_1992,krueger20}.
The possibility to construct easy and hard instances was one of the main motivations for selecting this problem for our experiments. Furthermore, the underlying decision problem (SAT) is considered the canonical example for NP-completeness~\cite{cook_complexity_1971} with numerous applications (for a broad overview see~\cite{kautz_state_2007,claessen_sat_2008}).
In our study, instances with \(\alpha \in (3.5, 4.9]\) are designated \enquote{hard}, and those outside this range \enquote{easy}. We evaluate ten instances for both, the \enquote{easy} and \enquote{hard} range, comprising 14--24 qubits, with one instance for each size.

The graphs and QUBOs of the individual instances used in the experiments can be found in our \repro.

\subsection{Experimental Setup}
\label{sec:exp-setup}
To eliminate detrimental influence of imperfections in actual quantum systems, all of our experiments, which evaluate the energy landscapes, were performed using noiseless, statevector simulations without finite sampling, with Qiskit~1.0.2~\cite{qiskit2024}. The results are fully reproducible~\cite{Mauerer:2022} via our \repro (link in PDF).
Additionally to the sequential method, we also explore the optimisation landscapes of linear-ramp (LR) parameter initialisation and a classical optimisation scheme, described in the following.

\subsubsection{LR-QAOA}
In the LR-QAOA scheme, parameters are set according to linear schedules, following Montañez-Barrera and Michielsen~\cite{montanez-barrera_towards_2024}. The ramps are parameterised by circuit depth \(p\) and slopes \((\Delta\beta,\Delta\gamma)\), corresponding to the mixer and phase operators, respectively. LR-QAOA is a well known method to efficiently produce high-quality results with approximation guarantees for various combinatorial optimisation problems \cite{montanez-barrera_towards_2024} at depths \(p>1\), without the need for extensive optimisation of the QAOA parameters.
Given differing implementations of mixers in the QAOA literature (\eg~Refs.~\cite{farhi_quantum_2014,Hadfield:2019,montanez-barrera_towards_2024}), we consider two cases:
(1) The signs of the slopes of both ramp parameters \(\Delta\beta,\Delta\gamma\) are positive, denoted as \(\LRp\), and the evolution uses \(U_B(\beta_j)=e^{-i\beta_iH_B}\) as defined in \autoref{sec:theory}.
(2) The sign of the slope of the mixer-ramp is negative, that is $-\Delta\beta,\Delta\gamma$, which is equivalent to using \(U_B(\beta_j)=e^{i\beta_iH_B}\) with a positive sign for the mixer ramp. We refer to this variant as as $\LRm$. The latter case matches the mixer used in Ref.~\cite{montanez-barrera_towards_2024}, likely as the ground state of this mixer corresponds to the initial state \(\ket{x}\), which in their work is chosen as \(\ket{x}=\ket{-}^{\otimes n}\). 
The purpose of this distinction is to examine how improper orientation of the ramps with regard to the initial state affects the results.
For both cases, we use $\Delta\gamma=0.6$ and $|\Delta\beta|=0.3$ as suggested by the authors in~\cite{montanez-barrera_towards_2024}.

\subsubsection{Classical Optimisation with COBYLA}
While we acknowledge that it may be infeasible to heavily optimise parameters in practical scenarios, in this work, we utilise the classical optimiser \COBYLA~\cite{powell_direct_1994} to provide optimised parameters as a comparison to the non-optimised methods.
COBYLA is a commonly used method for optimising QAOA parameters in the literature~\cite{campbell_qaoa_2022,montanezbarrera:2024,hao_end_2025} as it has the practical benefit of not requiring (expensive) gradient computations and therefore needing fewer function evaluations than gradient-based methods, while still achieving comparative results to other optimisers~\cite{pellow-jarman24}.
Initial values are obtained using sequential, \(\LRm\) and \(\LRp\) parameters for MaxCut, VertexCover and Max3SAT, respectively. These varied initialisations ensure sufficiently distinct starting points, allowing us to assess how the choice of initialisation influences  optimiser convergence and final parameter quality.
We terminate the optimisation process when the optimiser converges or the maximum function evaluations, set to $1000$, has been reached.

\subsubsection{Parameter Landscape Scans}
To probe the parameter landscape of LR and classical optimisation methods, we first obtain parameters for the target depth \(p_\text{target}\) using the method of interest.
Similar to the sequential method, for each \(p \in [1, p_\text{target}]\), parameters for depth \(p-1\) are fixed, while the parameters at level \(p\) are specified by a uniformly spaced grid.
However, unlike in the sequential method, these grid evaluations merely serve as an assessment tool and do not influence the method’s chosen parameters.
This procedure enables an organised assessment of parameter quality and allows comparison of parameters given by the subject initialisation methods with nearby alternatives in the optimisation landscape.
Further details on this landscape scan procedure are described in \autoref{sec:pseudocode}.

For all problem instances, we set \(p_\text{target} = 7\). Additionally, for MaxCut, we examine the \(p_\text{target} = 21\) landscapes for the same instances, incrementing \(p\) by 2 in each iteration. When optimising the \(p_\text{target} = 21\) instances, the parameters are initially set to the best values identified in the \LRp \(p=21\) landscape.

\section{Results}\label{sec:results}
We now discuss details of our numerical simulations, in particular with respect to properties of the parameter optimisation landscapes, as well as variations in and quality of results. Note that 
we postpone discussing interpretation and consequences to 
\autoref{sec:discussion}, and concentrate on the empirical
observations in the following.

\subsection{QAOA Parameter Symmetries} \label{subsec:symmetries}

\begin{figure*}[htb]
    \centering
    \includegraphics{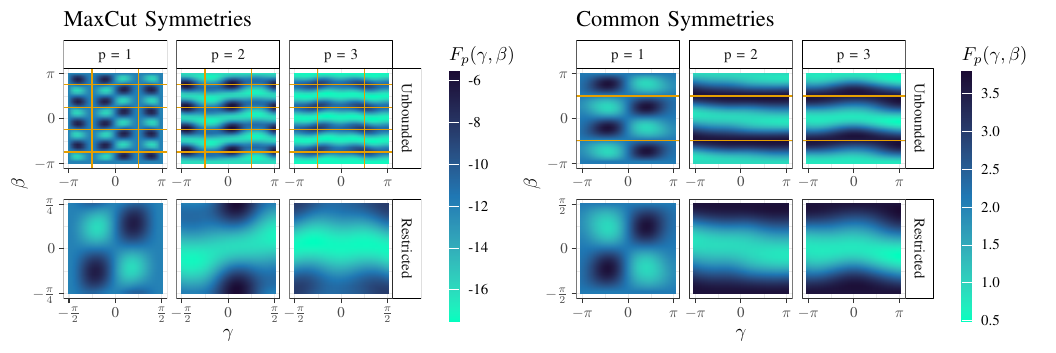}
    \caption{Symmetries specific to MaxCut (left) and common QAOA symmetries (right) illustrated by a single 16-vertex, 3-regular instance of MaxCut and a single 16 qubit Max3SAT instance with $\alpha=4.\overline{3}$. \emph{Top rows:} Overall optimisation landscape $F_{p}(\gamma, \beta)$ with $\gamma,\beta\in[-\pi,\pi]$ and symmetry axes marked in orange.
  \emph{Bottom rows:} Restriction to the subset $\gamma\in[-\frac{\pi}{2},\frac{\pi}{2}]$, $\beta\in[-\frac{\pi}{4},\frac{\pi}{4}]$ for MaxCut and $\gamma\in[-\pi,\pi]$, $\beta\in[-\frac{\pi}{2},\frac{\pi}{2}]$ in general that captures all information. The landscapes are obtained by the sequential method.}
    \label{fig:symmetries}
\end{figure*}

Our initial set of experiments aims to provide an intuitive visual understanding of symmetries that are inherent to the properties of QAOA (\cf~\autoref{sec:theory}), restrict the input domain and simplify numerical simulation.
Restricted symmetries for MaxCut on 3-regular graphs and general symmetries are illustrated in \autoref{fig:symmetries}.

The periodicity of the energy landscape can be observed in both parts of the figure, with period $\frac{\pi}{2}$ for $\beta$ and $\pi$ for $\gamma$, resulting in two horizontal identical sections in the case of common symmetries;
for the case of MaxCut this results in four horizontal and two vertical ($[-\frac{\pi}{2},\frac{\pi}{2}]$ range vs. $[\frac{\pi}{2},\pi]$ and  $[-\pi,-\frac{\pi}{2}]$ ranges) identical sections, respectively.
If we account for the symmetries and eliminate the corresponding degeneracies, the landscape can be restricted to the subset shown in the lower part of of the figure.

Additionally, QAOA parameters are invariant under time reversal, which is visualised by the fact that in the $p=1$ landscapes, quadrants $(-\beta,-\gamma$) and ($\beta,\gamma$),
as well as quadrants($-\beta,\gamma$) and ($\beta,-\gamma$) are point symmetric.
Notable, for $p>1$, time reversal symmetry may not always be visible in the plots, since we are only viewing a small part of the entire optimisation landscape given by all components of $\vec{\gamma},\vec{\beta}$.

\subsection{MaxCut}\label{subsec:maxcut}

\begin{figure*}[htb]
  \centering
  \includegraphics{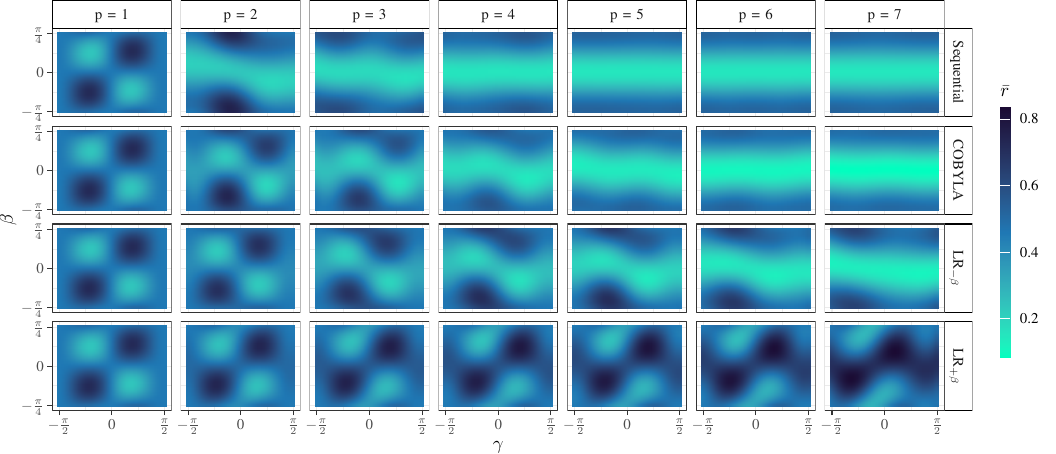}
  \caption{Average residual energy $\bar{r}$ for 40 3-regular MaxCut instances of sizes 10 to 16, with $\gamma,\beta$ set to sequentially fixed parameters (top row), optimised parameters using COBYLA starting from the sequential parameters (second row), $\LRm$ parameters with $\Delta_{\beta}=-0.3 , \Delta_{\gamma}=0.6$ (third row), and $\LRp$ parameters with $\Delta_{\beta}=0.3 , \Delta_{\gamma}=0.6$ (bottom row).}
  \label{fig:avg_parameter_scan_maxcut_symmetry}
\end{figure*}

\begin{figure*}[htb]
  \centering
  \includegraphics{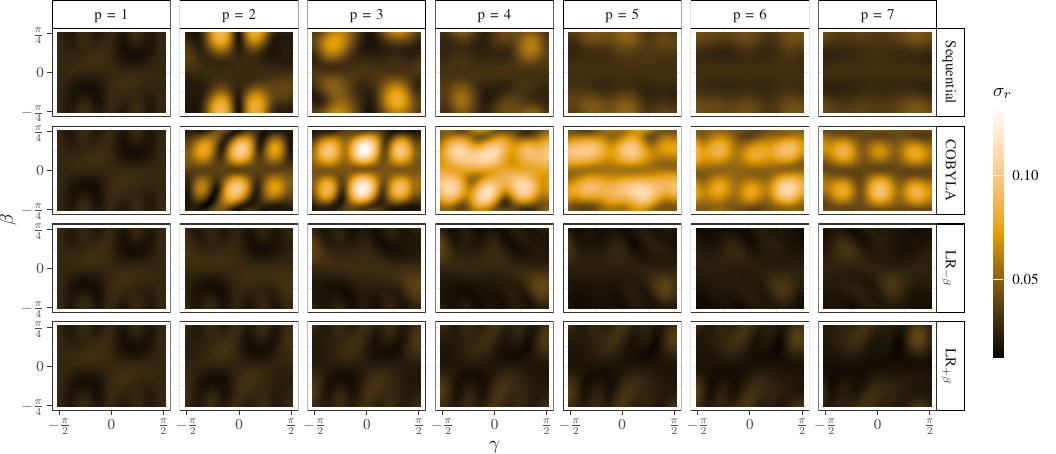}
  \caption{Standard deviation of the residual energy $\bar{r}$ in \autoref{fig:avg_parameter_scan_maxcut_symmetry} over 40 MaxCut instances.}
  \label{fig:std_parameter_scan_maxcut_symmetry}
\end{figure*}

\begin{figure}[]
  \centering
  \includegraphics{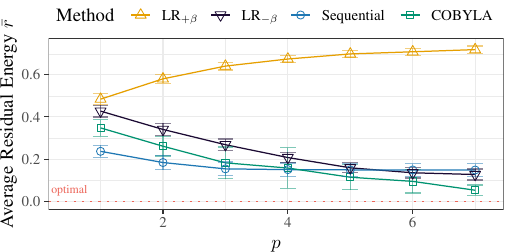}
  \caption{Average and standard deviation of the approximation quality for 40 3-regular MaxCut instances of sizes 10 to 16 when fixing parameters at each layer $p$ according to the considered methods. Note that since the COBLYA parameters are optimised for the target depth $p_\text{target}=7$ only, at lower depths performance might fall short. A reason for this behaviour is that the depth is insufficient to complete the full evolution of the optimised schedule, even performing worse than the sequential parameters, which were used to initialise the optimiser, at $p<4$.}
  \label{fig:params_parameter_scan_maxcut_symmetry} 
\end{figure}

Figures \ref{fig:avg_parameter_scan_maxcut_symmetry} and \ref{fig:std_parameter_scan_maxcut_symmetry} show average residual energy $\bar{r}$ and standard deviation $\sigma_r$ for our MaxCut instances.
Each plot facet corresponds to the optimisation landscape of $\vec{\beta}^{(p)},\vec{\gamma}^{(p)}$ with $\beta^{(p)}_{i},\gamma^{(p)}_{i} (0\leq i < p)$ components fixed to parameters of interest. Columns vary depth $p$, and different rows correspond to different parameter initialisation methods.
The average quality of the results produced by the parameters fixed at each layer by these methods is shown in \autoref{fig:params_parameter_scan_maxcut_symmetry}.
For comparison, the results of the smallest and largest individual instance are shown in \autoref{sec:maxcut_single}.

\paragraph{Average Residual Energy Landscapes}
The parameter landscapes obtained with the sequential method, COBYLA, and \(\LRm\) share common features: the optimal absolute value of \(\beta\) decreases with increasing \(p\), while the optimal absolute value of \(\gamma\), which grows at shallow depth, becomes effectively arbitrary at larger \(p\) once \(\beta \approx 0\) (see \autoref{fig:avg_parameter_scan_maxcut_symmetry}).  
In other words, the average residual energy landscapes given by the individual components of $\beta,\gamma$ become increasingly invariant under $\gamma$ as $\beta$ approaches zero.
This suggests that as \(\beta\) approaches zero, the system state is left unchanged by further temporal propagation under the cost Hamiltonian, while propagation under the mixer Hamiltonian still induces dynamics. Intuitively, if the system sufficiently approximates an energy eigenstate of the cost Hamiltonian, then adding an additional layer of cost and mixer operators leaves the state invariant under the cost Hamiltonian. In this regime, all choices of \(\gamma\) are therefore equally effective.  
Conversely, further propagation under the mixer Hamiltonian will change the state, so parameters $\beta$ need to be close to zero for the best results. In some circumstances, such as for the sequential method for $p>4$, the best choice of highest depth beta components is $\beta=0$, that is by not changing the state at all from the previous depth.
In this case, adding further layers with sequentially selected parameters will no longer yield improvements (and in fact in practical scenarios with noise, these additional layers may be detrimental to the results).
This therefore marks the point where adding of layers should be terminated for the sequential method.
In the further course of this paper we will often refer to this termination point as $\beta=0$, but this does not mean $\beta$ should actually be set to zero in real applications.
Accordingly, if the depth is sufficient for a QAOA circuit with a given set of parameters to adequately approximate one of the eigenstates of $H_C$, we can expect the landscape of the highest depth components of $\gamma,\beta$ to converge to such a landscape.
In contrast, the landscapes around the $\LRp$ parameters show the opposite of these convergence properties, with a choice of a small absolute value for $\beta$ leading to a \emph{high} $\bar{r}$, and the optimal choice of $\beta$ tends towards high values with increasing $p$.
This suggests that, rather than converging towards the ground state, the system approaches an excited state, which is natural, since a positive 
\(\Delta\beta\) corresponds to a positive mixer with 
\(\ket{+}\) as its maximally excited state. and with 
\(\Delta\gamma>0\), this corresponds to maximising \(H_C\).

The sequential method is comparatively the fastest to converge towards a landscape, where the energy is invariant under $\gamma$, which reaches such a state at a depth of $p=4$.
Fixing $\beta$ at (or very close to) zero (\cf \autoref{fig:parameter_arrangement}) at higher depths $p\geq4$ results in neither an increase nor decrease in $\bar{r}$ and no noticeable change in the landscapes.
When optimising the parameters using COBYLA, this transition between landscapes takes longer, reaching a landscape similar to the $p=4$ sequential landscape at $p=7$, with more intermediate transition steps.
The resulting $\bar{r}$ is higher than that of the sequential method at low depth ($p<4$), but starts outperforming the sequential method as the depth increases ($p>4$).
Unlike the sequential method, the quality keeps improving throughout all layers up to $p=7$ when optimising (see \autoref{fig:params_parameter_scan_maxcut_symmetry}).
The $\LRm$ parameters take the longest to converge to a state that is invariant under the phase operator, and in fact do not quite reach this point within $p=7$ with the given $\Delta\beta,\Delta\gamma$.
As shown in \autoref{fig:params_parameter_scan_maxcut_symmetry}, the average residual energy consistently improves at a similar rate to the optimised parameters, with the optimised parameters performing slightly better at all depths.
At $p\geq6$, The $\LRm$ parameters outperform the sequential parameters.
The average residual energy given by the $\LRp$ parameters changes inversely to the $\LRm$, albeit the landscapes of these methods are not an inverse of each other, with the individual $\LRp$ landscapes differing less from preceding ones than for the $\LRm$ parameters.

\paragraph{Convergence to $\gamma$ Invariance at $\beta=0$}

Another notable observation from \autoref{fig:avg_parameter_scan_maxcut_symmetry} is that the maxima seem to shift in accordance with the changes in $\gamma$ and $\beta$ with increasing $p$:
(1) the maxima shift towards the top/bottom of the landscape (\ie, the symmetry boundary of $\beta$) and gradually diminish, while minima shift towards $\beta=0$;
and (2) the maximal/minimal regions, which are duplicated due to time-reversal symmetry, broaden diagonally and merge into a single maximal/minimal region.

For  $\LRm$, higher depths correspond to a high $\gamma$ relative to $\beta$ and lower depths to a low $\gamma$ relative to $\beta$. Comparing theses ranges illustrates that the minima/maxima shifts are consistent with the changes in $\beta,\gamma$:
At low depth ($p<4$), where $\gamma$ is (near) zero, property (2) is more pronounced than property (1), with the position of the extrema barely changing.
Only the width of the minimal/maximal region increases along the diagonal between duplicate minima/maxima.
At higher depths ($p>4$), as $\gamma$ increases (and $\beta$ approaches zero), property (1) becomes more pronounced.
The sequential method fixes non-zero values for both $\beta$ and $\gamma$ at $p=1$ and subsequently, at $p=2$, reaches a landscape similar to the $p=6$ $\LRm$ parameters  showing both properties (1) and (2).
The lower depth ($p<4$) scans show that the COBYLA optimiser has a tendency to chose higher values of $\beta$ and lower values of $\gamma$ than the sequential method (\cf~\autoref{fig:parameter_arrangement}).
Furthermore, property (2) is more pronounced for the sequential parameters than those obtained with COBYLA, which at low depth have a longer propagation under $\beta$ relative to the chosen $\gamma$.
This indicates that for COBYLA property (1) becomes more pronounced as $\beta$ decreases, relative to $\gamma$.

\paragraph{Residual Energy Standard Deviation Landscapes}

At $p=1$, the standard deviation (see \autoref{fig:std_parameter_scan_maxcut_symmetry}) from $\bar{r}$ is low ($\sigma_r\leq 0.05$) across the entire landscape, with areas of almost zero $\sigma_r$ in parts of the landscape where $\beta\approx\pm\frac{\pi}{8},\gamma\approx\pm\frac{\pi}{4}$.
The parts of the landscape which correspond to the minima in the $p=1$ $\bar{r}$ landscape in \autoref{fig:avg_parameter_scan_maxcut_symmetry} result in $\sigma_r$ not significantly different from the rest of the landscape ($\leq 0.05$).
At $p=2$ to $p=4$, for the sequential method, there are some areas of moderate standard deviation ($\sigma_r\leq 0.1$) that correspond to a long propagation under $\beta\approx\pm\frac{\pi}{4}$ and specific values of $\gamma$.
In the corresponding $\bar{r}$ landscapes of \autoref{fig:avg_parameter_scan_maxcut_symmetry}, these regions of moderate $\sigma_r$ contain maxima or medium $\bar{r}$.
At $p>4$, $\sigma_r$ is more uniform across the entire landscape, with the overall magnitude $\sigma_r$ decreasing with increasing $p$, but with higher $\sigma_r$ as in the $p=1$ landscape.
When optimising the sequential parameters using COBYLA, $\sigma_r$ is comparatively higher ($0.1\leq\sigma_r\leq0.15$ in some areas of the landscapes between $p=3$ and $p=6$) and less uniformly distributed.
There are some narrow regions where $\sigma_r$ is lower than the rest of the landscape, namely in the areas where $\beta=0$ or $\beta=\pm\frac{\pi}{4}$ and in areas where $\gamma\approx\pm\frac{\pi}{4}$ (the exact $\gamma$ seems to vary slightly), making the landscape appear segmented into areas of high $\sigma_r$.
Both \LR parameters result in landscapes with low, uniform $\sigma_r$ that barely differ from the $p=1$ landscape, with $\sigma_r$ only slightly decreasing in some areas and slightly increasing in others.
For the $\LRm$ parameters, the areas in the middle of the landscape increase in $\sigma_r$, while for the $\LRp$ parameters the $\sigma_r$  decreases in these areas.
For the areas at the edges of the plot, the inverse of this applies.

\subsection{MaxCut: Higher depths}\label{subsec:higher_p}

\begin{figure*}[htb]
  \centering
  \includegraphics{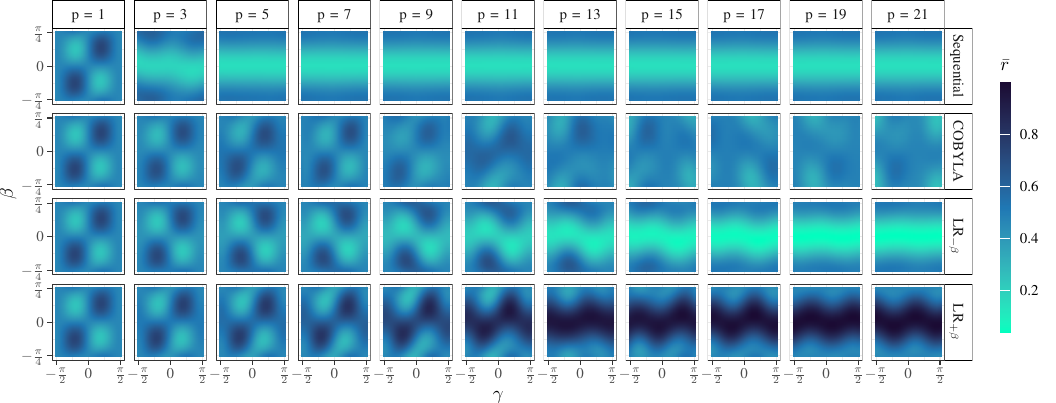}
  \caption{Average residual energy $\bar{r}$ for the same 40 MaxCut instances of sizes 10 to 16 shown in \autoref{fig:avg_parameter_scan_maxcut_symmetry} with $p_\text{target}=21$, with $\gamma,\beta$ set to sequentially fixed parameters (top row), optimised parameters using COBYLA starting from the $\LRp$ parameters (second row), $\LRm$ parameters with $\Delta_{\beta}=-0.3 , \Delta_{\gamma}=0.6$ (third row), and $\LRp$ parameters with $\Delta_{\beta}=0.3 , \Delta_{\gamma}=0.6$ (bottom row).}
  \label{fig:avg_parameter_scan_higher_p_symmetry}
\end{figure*}

\begin{figure*}[htb]
  \centering
  \includegraphics{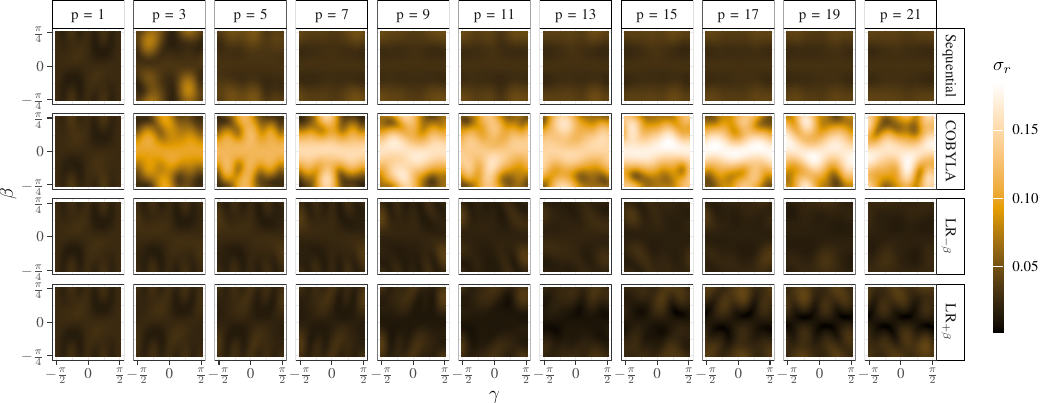}
  \caption{Standard deviation of the residual energy $\bar{r}$ in \autoref{fig:avg_parameter_scan_higher_p_symmetry} over 40 MaxCut instances, where $p_\text{target}=21$.}
  \label{fig:std_parameter_scan_higher_p_symmetry}
\end{figure*}

\begin{figure}[h]
  \centering
  \includegraphics{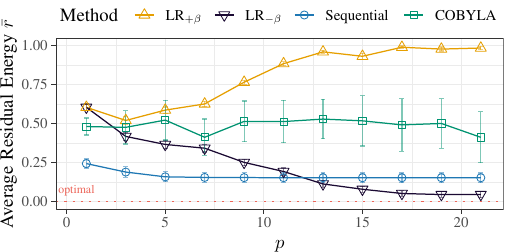}
  \caption{Average and standard deviation of the approximation quality for 40 3-regular MaxCut instances of sizes 10 to 16 when fixing parameters at each layer $p$ according to the considered methods, where $p_\text{target}=21$.}
  \label{fig:params_parameter_scan_higher_p_symmetry}
\end{figure}

To examine the parameter landscapes at higher depths we further evaluate the same MaxCut instances up to $p_\text{target}=21$.
The results of these experiments are shown in Figures \ref{fig:avg_parameter_scan_higher_p_symmetry} and \ref{fig:std_parameter_scan_higher_p_symmetry}, respectively corresponding to the average residual energy $\bar{r}$ and the standard deviation $\sigma_r$.
The landscape is evaluated in intervals of two layers for COBYLA and the \LR parameters, omitting every second landscape, to save computation time due to the increased number of circuit evaluations and circuit depths.
The average quality of the energies produced by the parameters fixed at each layer by these methods is shown in \autoref{fig:params_parameter_scan_higher_p_symmetry}. 
The results for one individual instance are shown in \autoref{sec:maxcut_single}.

\paragraph{Average Residual Energy Landscapes}

At higher depths, the landscape around the sequential parameters (top row of \autoref{fig:avg_parameter_scan_higher_p_symmetry}) does not change in a noticeable way compared to the $p_\text{target}=7$ landscape, with the lowest energy values occurring at $\beta=0$ and arbitrary $\gamma$.
As with the lower depth experiments (\autoref{subsec:maxcut}), from $p=4$ onwards, the sequential method consistently fixes $\beta=0$ (\cf~\autoref{fig:parameter_arrangement}), that is, no further temporal propagation under the mixer occurs at higher depths, and since any propagation under the phase operator leaves the system in its state, the residual energy also remains constant (\cf~\autoref{fig:params_parameter_scan_higher_p_symmetry}).
This behaviour aligns with our expectations, since the sequential method always chooses the best available parameters at each $p$ without modifying previously fixed parameters.
So, once it has reached a landscape where further propagation under $\beta$ only results in higher energies and the choice of $\gamma$ is arbitrary, there is no further propagation under the mixer at higher depths.

The landscapes around the $\LRm$ parameters (lower-middle row of \autoref{fig:avg_parameter_scan_higher_p_symmetry}) progress similarly to the landscapes in the lower depth experiments, but at an overall slower rate.
This is natural, as the slope of the linear ramp is $\frac{\Delta}{p}$, and using a higher $p$ with the same $\Delta$ corresponds to a more gradual slope, which in turn corresponds to more gradual changes in the variational parameters.
The higher depth landscape at $p>7$ continues to converge to a landscape that is dependant on $\beta$ and invariant under $\gamma$, and unlike the lower depth case, there is no more variation in $\bar{r}$ depending on $\gamma$ at $p=21$. 
The performance of the higher depth $\LRm$ parameters (see \autoref{fig:params_parameter_scan_higher_p_symmetry}) at low depth $p<13$ is worse than the low depth $\LRm$ parameters (compare \autoref{fig:params_parameter_scan_maxcut_symmetry}), but after $p=13$, the higher depth $\LRm$ parameters start outperforming the lower depth variant, as well as the sequential parameters.
At $p\geq 17$ a lower $\bar{r}$ can be achieved than the best-performing optimised parameters at lower depth (\cf~\autoref{fig:params_parameter_scan_maxcut_symmetry}).

The $\LRp$ landscape (bottom row of \autoref{fig:avg_parameter_scan_higher_p_symmetry}) similarly progresses at a slower rate compared to to the respective lower depth landscapes.
The final landscape is invariant under $\gamma$ close to $\beta=0$, albeit there is some variation in $\bar{r}$ close to $\beta=\pm\frac{\pi}{4}$ depending on $\gamma$.

When using COBYLA to optimise the $\LRp$ parameters, the resulting landscapes are distinct from the case, where we initialised the optimiser with sequential parameters at $p_\text{target}=7$.
The landscapes mainly consist of rather average $\bar{r}$ and the maxima and minima of the individual landscapes are less pronounced, with convergence patterns becoming less discernable as $p$ increases.
At $p \leq 11$, the landscapes progress similarly to the $\LRp$ landscapes.
However at $p > 11$, the maxima become less pronounced, compared to the the $\LRp$ parameters, and the minima shift horizontally towards the corners of the landscape at $p = 21$, which corresponds to a long propagation under both $\beta$ and $\gamma$.
Overall, the optimiser performs worse than the sequential parameters and the $\LRm$ parameters (see \autoref{fig:params_parameter_scan_higher_p_symmetry}), as well as the optimised sequential parameters in the lower depth case (\cf~\autoref{fig:params_parameter_scan_maxcut_symmetry}).

\paragraph{Residual Energy Standard Deviation Landscapes}

The standard deviation of these landscapes (see \autoref{fig:std_parameter_scan_higher_p_symmetry}) at higher depths behaves similarly to the $p_\text{target}=7$ experiments:
The sequential parameters and the $\LRm$ parameters both result in a low standard deviation across the entire landscape, with $\sigma_{r}<0.1$ and $\sigma_r\leq 0.05$, respectively.
$\LRp$ parameters also result in a similar $\sigma_r$ as the lower depth case (\cf \autoref{fig:std_parameter_scan_maxcut_symmetry}), but with slightly higher $\sigma_r<0.1$ occurring in the landscape where $\beta$ is non-zero.
The landscapes of COBYLA initialised with $\LRp$ parameters have the highest $\sigma_r<0.2$, across a large part of the landscape, with the highest $\sigma_r$ generally occurring in the part of the landscape that corresponds to a short propagation under the mixer, that is, where $\beta$ is close to zero.
This indicates that there is a relatively high variation in the cost landscape across MaxCut instances when optimising $\LRp$ parameters.
Notably, there are no areas of low $\sigma_r$ at specific values of $\gamma$ or $\beta$, such that the $\sigma_r$ landscape is divided into segments, as was observed in the lower depth experiments, where COBYLA was initialised with the sequential parameters (see \autoref{subsec:maxcut}).

\subsection{VertexCover}\label{subsec:vertexcover}

\begin{figure*}[htb]
  \centering
  \includegraphics{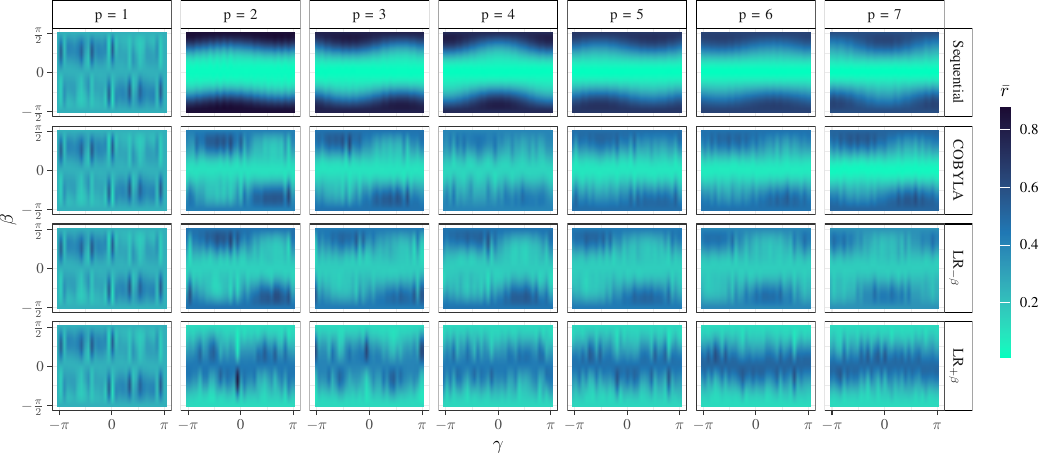}
  \caption{Average residual energy $\bar{r}$ for 40 VertexCover instances of sizes 10 to 16, with $\gamma,\beta$ set to sequentially fixed parameters (top row), optimised parameters using COBYLA starting from the $\LRm$ parameters (second row), $\LRm$ parameters with $\Delta_{\beta}=-0.3 , \Delta_{\gamma}=0.6$ (third row), and $\LRp$ parameters with $\Delta_{\beta}=0.3 , \Delta_{\gamma}=0.6$ (bottom row).}
  \label{fig:avg_parameter_scan_vertcover_symmetry}
\end{figure*}

\begin{figure*}[htb]
  \centering
  \includegraphics{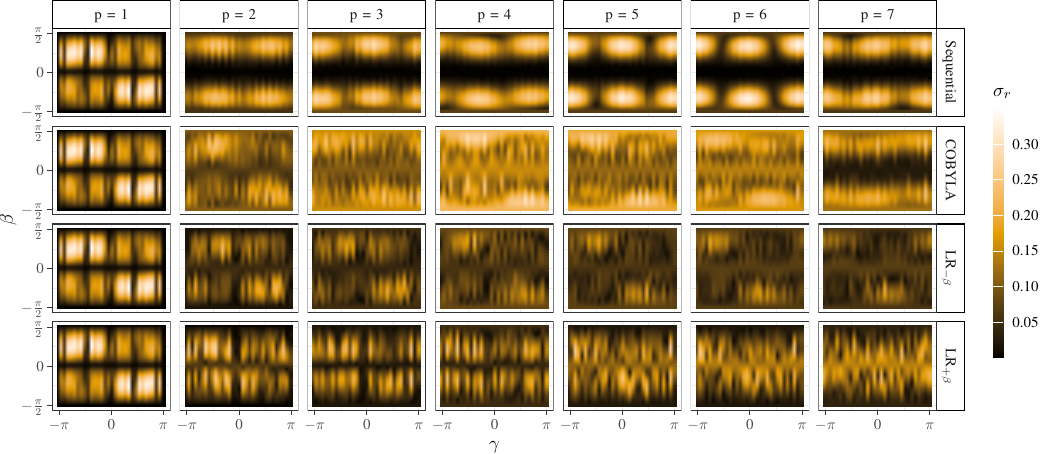}
  \caption{Standard deviation of the residual energy $\bar{r}$ in \autoref{fig:avg_parameter_scan_vertcover_symmetry} over 40 VertexCover instances.}
  \label{fig:std_parameter_scan_vertcover_symmetry}
\end{figure*}

\begin{figure}[]
  \centering
  \includegraphics{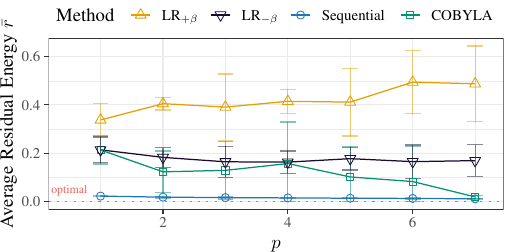}
  \caption{Average and standard deviation of the approximation quality for 40 VertexCover instances of sizes 10 to 16 when fixing parameters at each layer $p$ according to the considered methods.}
  \label{fig:params_parameter_scan_vertcover_symmetry}
\end{figure}

Figures \ref{fig:avg_parameter_scan_vertcover_symmetry} and \ref{fig:std_parameter_scan_vertcover_symmetry} respectively show the average residual energy $\bar{r}$ and the standard deviation $\sigma_r$ of the VertexCover instances on the 40 3-regular graph instances used in Sections \ref{subsec:maxcut} and \ref{subsec:higher_p}, up to $p_\text{target}=7$.
The average quality of the energies produced by the parameters using the described methods is shown in \autoref{fig:params_parameter_scan_vertcover_symmetry}. 
The results for two single instances (smallest and largest) are shown in \autoref{sec:vertexcover_single}.
Aside from the fact that here we initialise COBYLA with $\LRm$ parameters, the figures use the same methods and follow the same structures as the MaxCut experiments (\cf~\autoref{subsec:maxcut}).

\paragraph{Average Residual Energy Landscapes}
In comparison to MaxCut, the $p=1$ landscape for VertexCover exhibits a higher density of local minima and maxima.
A possible reason for this is that there are more computational basis states corresponding to invalid solutions for VertexCover than for MaxCut.
A slight variation to the parameters may cause a transition from a valid solution state to an invalid solution state, leading to irregularities in the surface of the optimisation landscape.
Furthermore, as discussed in \autoref{subsec:symmetries}, the range in which $\gamma,\beta$ are periodic is different to MaxCut, as only general symmetries apply for VertexCover.
Nevertheless, the $p=1$ VertexCover landscape shares some general properties with the $p=1$ MaxCut landscape, such as time-reversal symmetry.
Additionally, the landscape follows the general structure, with the lower left and upper right quadrants in general containing lower values of $\bar{r}$ than the lower right and upper left quadrants.

As $p$ increases, the sequential method quickly converges to a landscape with a smoother surface, where $\bar{r}$ is invariant under $\gamma$ when $\beta=0$ or $\beta=\pm\frac{\pi}{2}$, at $p\geq 2$.
However, unlike in the MaxCut experiments, there is no examined depth, where the landscape is entirely invariant under $\gamma$ for any choice of $\beta$.
From $p\geq3$ onwards, the sequential method fixes $\beta$ close to zero (\cf~\autoref{fig:parameter_arrangement}) and $\bar{r}$ remains constant (\cf~\autoref{fig:params_parameter_scan_vertcover_symmetry}), but the location and magnitude of the maxima at the edges of the landscape still change depending on the chosen $\gamma$.
This suggests that the system is in a state, in which further propagation under the phase operator will leave the system in the state, if there is no propagation under the mixer.
Evolution under the mixer in turn may change the state depending on the magnitude of both $\gamma$ and $\beta$, indicating that the propagated state no longer closely approximates an eigenstate of $H_C$.
In short, the effect of propagation under $\beta$ is more pronounced than for the MaxCut experiments (see \autoref{subsec:maxcut} and \autoref{subsec:higher_p}).

In the $\LRm$ landscape (third row of \autoref{fig:avg_parameter_scan_vertcover_symmetry}) at $p \geq 2$ the minima also shift towards the center of the plot, but in comparison to the landscape of the sequential method, it is overall less smooth and the minima and maxima are less extensive.
Furthermore, while the parts of the landscape where $\beta=0$ contain similarly low values of $\bar{r}$, the landscape is never completely invariant under $\gamma$ in these regions, suggesting that the $\LRm$ fail to converge to a state which closely approximates an eigenstate of $H_C$.
This landscape barely changes at higher depths $p\geq3$.

When optimising the $\LRm$ parameters using COBYLA (upper-middle row of \autoref{fig:avg_parameter_scan_vertcover_symmetry}), the $p=2$ and $p=3$ landscapes are similar to the $\LRm$ landscapes for the same $p$.
However, as $p$ increases further, the landscapes converge to a state that is invariant under $\gamma$ at $\beta=0$.
Compared to the landscapes of the sequential method, the optimiser landscapes are less smooth and the minimum is narrower in $\beta$, with less pronounced maxima up to $p=7$.
\autoref{fig:params_parameter_scan_vertcover_symmetry} also shows a similar performance of the COBYLA and sequential method at $p=7$.

The $\LRp$ parameters (bottom row of \autoref{fig:avg_parameter_scan_vertcover_symmetry}), converge towards the inverse of the other landscapes, and similar to the $\LRm$, do not reach a landscape which is invariant under $\gamma$ when $\beta=0$, instead only reaching a landscape with low variance under $\gamma$ at $\beta=0$.
Unlike the $\LRm$ parameters, the $\bar{r}$ of the $\LRp$ keeps changing with increasing $p$, towards higher values of $\bar{r}$ (see \autoref{fig:params_parameter_scan_vertcover_symmetry}).
Notably, at $p=2$ with $\gamma$ slightly below zero, there is a relatively high maximum for negative $\beta$ and a relatively low minimum for positive $\beta$.
These extrema are mirrored in $\beta$ at $p=3$, suggesting that some specific combinations of $\gamma,\beta$ at varying depth lead to opposite behaviours.

\paragraph{Convergence to $\gamma$ Invariance at $\beta=0$}

As with MaxCut, the maxima and minima in the landscapes seem to shift and broaden depending on the magnitude of $\gamma,\beta$, although this is less evident due to the dense local minima/maxima in the landscape.
The most notable example here is evident in the sequential method, for which $\beta$ quickly decreases to zero within $p\leq3$ (\cf \autoref{fig:parameter_arrangement}), as the maximal/minimal regions quickly merge accordingly.
As the state is further propagated under $\gamma$, the maxima become diminished and tend towards the vertical extremes of the landscape while the minima become narrower around $\beta=0$.

\paragraph{Residual Energy Standard Deviation Landscapes}

The standard deviation landscape (\cf~\autoref{fig:std_parameter_scan_vertcover_symmetry}) at $p=1$ shows a relatively high $\sigma_r$ in the upper left and lower right quadrants, which tend to contain higher values of $\bar{r}$ in the average landscapes (\cf~\autoref{fig:avg_parameter_scan_vertcover_symmetry}).
In the other quadrants, which tend to contain lower $\bar{r}$, $\sigma_r$ is medium in magnitude.
In the areas, in which $\beta=0$ and $\beta=\pm\frac{\pi}{2}$ with medium $\bar{r}$, or in areas with pronounced maxima and minima of $\bar{r}$, $\sigma_r$ is close to zero.
At higher depths, for the sequential method $\sigma_r$ is close to zero in the regions where $\beta$ is (near) zero, and $\bar{r}$ is minimal.
The standard deviation is higher and more variable in the areas where $\beta$ is closer to $\pm\frac{\pi}{2}$ and $\bar{r}$ is higher.
Notably, the landscapes, which the sequential method converges against, lie in areas of low $\sigma_r$ close to $\gamma=\pm\frac{\pi}{2}$ for $p=5$ and $p=6$, though this effect is less recognizable at $p=7$.
The \LR parameters result in landscapes where middling and higher values of $\sigma_r$ are distributed more uniformly across the landscape, with frequent narrow areas of low standard deviation occurring in irregular intervals of $\gamma$.
In general, for the $\LRm$ parameters, higher $\sigma_{r}$ lie in areas corresponding to a high $\bar{r}$, and more middling $\sigma_r$ in areas with low $\bar{r}$.
The areas with low $\sigma_r$ are rather uneven, partly corresponding to more pronounced extrema and partly to a medium $\bar{r}$.
The $\LRp$ parameters result in landscapes where areas of low $\sigma_r$ seem to appear more regularly at low $p$ near $\beta=0$ and in areas with pronounced extrema of $\bar{r}$, as well as at high $p$, in areas of the where $\beta=\pm\frac{\pi}{2}$.
However, overall high values of $\sigma_r$ tend to occur more frequently and to a greater extent than in the $\LRm$ landscapes.
When optimising the $\LRm$ parameters with COBYLA, the standard deviation of the resulting landscapes increases noticeably and is spread more uniformly across the entire landscape as $p$ increases.
Notably, after fixing $p=6$ optimised parameters, a transition occurs in the $\sigma_r$ landscape, causing the $p=7$ landscape to be more similar to that of the sequential parameters at $p=7$, with low $\sigma_r$ in the region where $\beta$ is near 0, and higher $\sigma_r$ where $\beta$ is closer to $\pm\frac{\pi}{2}$.

\subsection{Max3SAT}\label{subsec:max3sat}

\begin{figure*}[htb]
  \centering
  \includegraphics{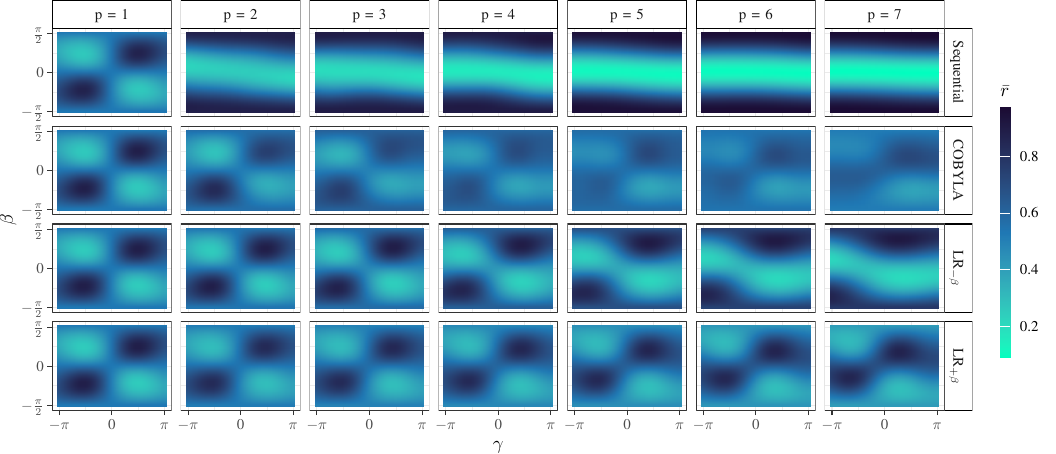}
  \caption{Average residual energy $\bar{r}$ for 10 Max3SAT instances of sizes 14 to 24, with $\alpha\in(3.5;4.9]$ and $\gamma,\beta$ set to sequentially fixed parameters (top row), optimised parameters using COBYLA starting from the $\LRp$ parameters (second row), $\LRm$ parameters with $\Delta_{\beta}=-0.3 , \Delta_{\gamma}=0.6$ (third row), and $\LRp$ parameters with $\Delta_{\beta}=0.3 , \Delta_{\gamma}=0.6$ (bottom row).}
  \label{fig:avg_parameter_scan_max3sat_hard_symmetry}
\end{figure*}

\begin{figure*}[htb]
  \centering
  \includegraphics{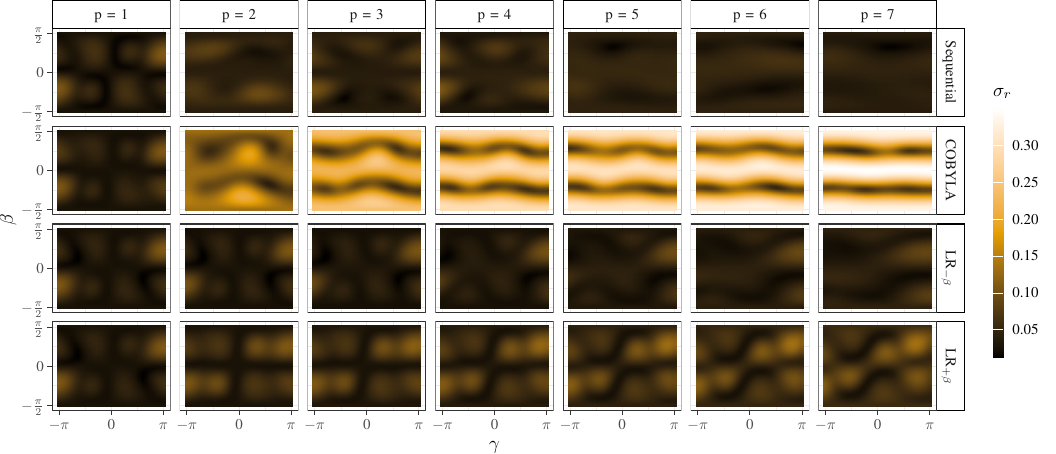}
  \caption{Standard deviation of the residual energy $\bar{r}$ in \autoref{fig:avg_parameter_scan_max3sat_hard_symmetry} over 10 Max3SAT instances.}
  \label{fig:std_parameter_scan_max3sat_hard_symmetry}
\end{figure*}

Figures \ref{fig:avg_parameter_scan_max3sat_hard_symmetry} and \ref{fig:std_parameter_scan_max3sat_hard_symmetry} show the average residual energy $\bar{r}$ and the standard deviation for 10 \enquote{hard} Max3SAT instances with $\alpha\in(3.5;4.9]$~\cite{mitchell_hard_1992}, up to $p_\text{target}=7$.
The average quality of the energies produced by the parameters fixed at each layer is shown in \autoref{fig:params_parameter_scan_vertcover_symmetry}.
The results for two single instance are shown in \autoref{sec:max3sat_hard_single}.
Apart from initialising the COBYLA optimiser with $\LRp$ parameters, the figures use the same methods and follow the same structures as the previous experiments (\cf~\autoref{subsec:maxcut} or \autoref{subsec:vertexcover}).
Additionally, we evaluated 10 Max3SAT instances from the easy range, $\alpha\notin(3.5;4.9]$, which gave qualitatively similar results, as we describe in \autoref{sec:max3sat_easy}.

\begin{figure}[h]
  \centering
  \includegraphics{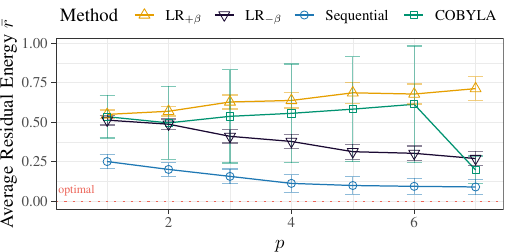}
  \caption{Average and standard deviation of the approximation quality for 10 Max3SAT instances of sizes 14 to 24 with $\alpha\in(3.5;4.9]$ when fixing parameters at each layer $p$ according to the considered methods.}
  \label{fig:params_parameter_scan_max3sat_hard_symmetry}
\end{figure}

\paragraph{Average Residual Energy Landscapes}

The resulting landscapes resemble those of MaxCut, but exhibit key differences:
As discussed in \autoref{subsec:symmetries}, the MaxCut specific symmetries do not apply for Max3SAT; only the general $\beta$-symmetry and time reversal symmetry applies.
Compared to MaxCut, the Max3SAT landscapes contain fewer regions of intermediate $\bar{r}$, dominated instead by high and low values.
At higher $p$, the maxima are larger, whereas for MaxCut the highest $\bar{r}$ decrease with depth (\cf~\autoref{subsec:maxcut}).
Overall, all methods perform worse than in previous experiments, generally yielding higher (or lower for $\LRp$) $\bar{r}$ (\cf~\autoref{fig:params_parameter_scan_max3sat_hard_symmetry} and \autoref{fig:params_parameter_scan_maxcut_symmetry}, also \cf~\autoref{fig:params_parameter_scan_higher_p_symmetry} and \autoref{fig:params_parameter_scan_vertcover_symmetry}).
This suggests that for Max3SAT, inadequate temporal propagation (\ie, the system evolves away from the target state) at higher depth result in higher $\bar{r}$ than for inadequately propagated QAOA in MaxCut and Vertexcover at the same depths.
The sequential method landscape (top row of \autoref{fig:max3sat_landscape}) progresses similarly to prior experiments, approaching a $\gamma$-invariant and $\beta$-periodic structure, with minima at $\beta=0$ and maxima at $\beta=\pm\frac{\pi}{2}$.
Unlike other cases, the sequential method's average $\bar{r}$ continues to decrease up to $p=7$, though the improvement becomes negligible beyond $p=5$ (\cf~\autoref{fig:params_parameter_scan_max3sat_hard_symmetry}).
Analogously, the \LR landscapes evolve more slowly, showing intermediate $\bar{r}$ as $p$ increases.
At $p=7$, neither the $\LRm$ nor the $\LRp$ have converged to $\gamma$-invariance at $\beta=0$:
For $\LRm$, the best parameters occur near $\beta\approx\frac{\pi}{8}$, $\gamma\approx-\pi$ or $\beta\approx-\frac{\pi}{8}$, $\gamma\approx\frac{\pi}{3}$; for $\LRp$ the extrema appear at the corresponding inverted-sign regions.
The COBYLA optimiser, initialised with the $\LRp$ parameters, produces increasingly, but intermediate $\bar{r}$, and weaker extrema as $p$ increases.

None of the methods closely approximates the true ground state energy (\cf~\autoref{fig:params_parameter_scan_max3sat_hard_symmetry}).
The sequential method performs best, saturating beyond $p=5$;
$\LRm$ ranks second for $p < 7$, improving comparatively to the sequential method up to $p = 4$ but starting from a worse $\bar{r}$ at $p=1$.
COBYLA, initialised with energy maximising $\LRp$ parameters, performs similarly to $\LRm$ at $p\leq2$, worse from $p=3$ to $p=6$, and substantially better at $p=7$, surpassing $\LRm$, but performing worse than the sequential method.

\paragraph{Convergence to $\gamma$ Invariance at $\beta=0$}

As in \autoref{subsec:maxcut}, \ref{subsec:higher_p} and \ref{subsec:vertexcover}, landscape extrema shift and broaden with increasing $\gamma$ and decreasing $\beta$.
The \LR parameters illustrate this gradual transition:
At $p=2$, for high $\beta$ and small $\gamma$, the landscape resembles that of $p=1$.
With increasing $p$, duplicate extrema broaden diagonally and merge.
Whether minima or maxima merge across the central diagonal depends on the $\beta$ ramp sign: minima merge for negative $\beta$, maxima for positive $\beta$.
For $p\geq4$, as $\beta \to 0$ and $\gamma$ increases, extrema shift vertically, with minima moving inward and maxima outward for negative $\beta$, with the reverse for positive $\beta$.
These shifts intensify as $\gamma$ approaches $\Delta_{\gamma}$.

\paragraph{Residual Energy Standard Deviation Landscapes}

At $p=1$, the standard deviation (\cf~\autoref{fig:std_parameter_scan_max3sat_hard_symmetry}) is generally relative low, but slightly higher than for $p=1$ MaxCut (\cf~\autoref{subsec:maxcut},\ref{subsec:higher_p}) and lower than for $p=1$ VertexCover (\cf~\ref{subsec:vertexcover}).
The largest $\sigma_r$ appear near $\gamma=\pm\pi$ and $\beta=\pm\frac{\pi}{4}$, coinciding with the extrema of the $p=1$ $\bar{r}$ landscape (\cf~\autoref{fig:avg_parameter_scan_max3sat_hard_symmetry}).
Regions near $\beta=0$ or $\gamma=0$ show nearly zero $\sigma_r$, segmenting the landscape into high- and low-variance zones.
As $p$ increases, the sequential method's landscape becomes more uniform: peak $\sigma_r$ remains stable up to $p=4$ and decreases slightly from $p\geq5$.
The \LR landscapes, homogenise more slowly; the $\LRm$ landscapes change minimally up to $p=3$, while $\LRp$ landscapes transition slightly faster at $p=2$, but then plateau.
Overall, $\LRp$ exhibits higher $\sigma_r$ than $\LRp$.
For COBYLA initialised with $\LRp$ parameters, higher-depth ($p\geq 2$) landscapes contain high $\sigma_r$, approaching $\approx0.35$ as $p$ increases.
Low-variance regions ($\sigma_r\leq0.1$) near $\beta=\pm\frac{\pi}{4}$ become progressively less dependant on $\gamma$, so that by $p=7$, $\sigma_r$ varies little with $\gamma$.
This suggests that higher-depth runs with $\beta=\pm\frac{\pi}{4}$ yield $r$-values closely clustered around $\bar{r}$.
However, this does not necessarily indicate that an eigenstate of $H_C$ is being approximated as $\bar{r}$ is still $\gamma$-dependant.

\subsection{(Non-)Patterns in QAOA Parameters}

\begin{figure*}[htb]
  \centering
  \includegraphics{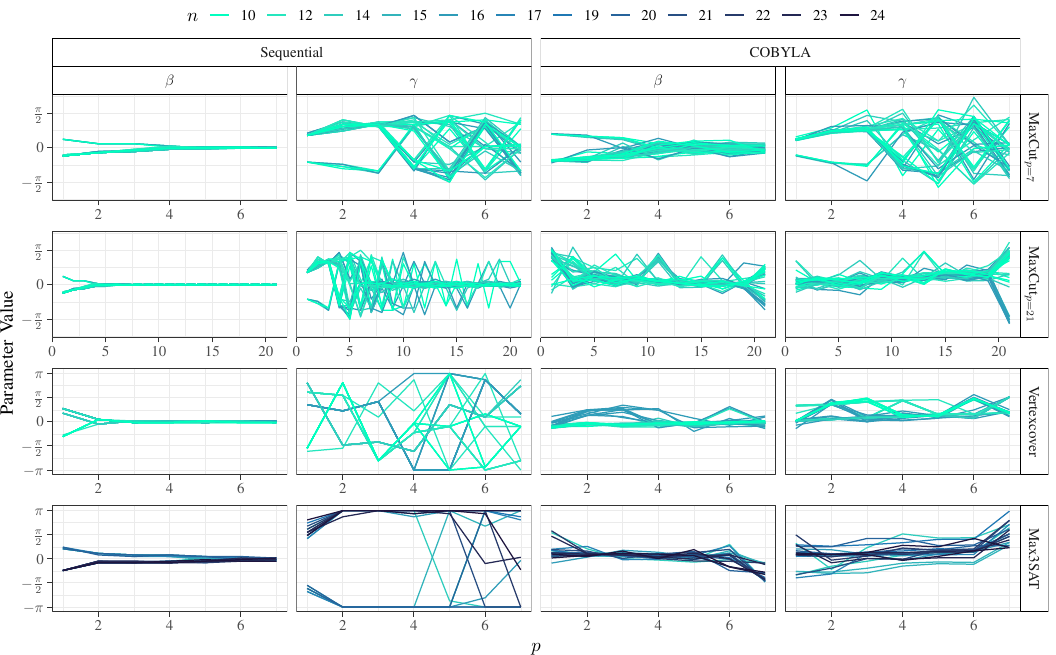}
  \caption{Parameter components of $\vec{\gamma},\vec{\beta}$ at each $p$ determined by our sequential method and the COBYLA optimiser for each instance of the considered optimisation problems. Colour denotes number of qubits $n$ for each problem instance. For each problem different parameters were used to initialise COBYLA: sequential parameters for $p_\text{target}=7$ MaxCut (top row), \LRp parameters for $p_\text{target}=21$ MaxCut (second row) and Max3Sat (bottom row), and \LRm parameters for VertexCover (third row).}
  \label{fig:parameter_arrangement}
\end{figure*}

To assess how well the examined parameters conform to the patterns described in the literature, \autoref{fig:parameter_arrangement} shows the parameters determined by the sequential method (left) and COBYLA (right).
The \LR ramps are omitted, as they follow a linear pattern by construction.

For the sequential method, the $\beta$ parameters follow the expected pattern, decreasing (non-linearly), rapidly at low $p$ and gradually at higher $p$.
The $\gamma$ parameters adhere to literature trends only for low depth ($p\leq3$ for MaxCut, $p\leq2$ for Max3SAT).
At higher depth and for $p\geq1$ VertexCover, $\gamma$ varies irregularly, often fixed to $\pm\pi$ for larger $p\geq2$ Max3SAT instances.
These deviations likely arise because the sequential method quickly converges to a $\gamma$-invariant state when $\beta=0$ (\cf~\autoref{subsec:maxcut}, \ref{subsec:higher_p}, \ref{subsec:vertexcover} and \ref{subsec:max3sat}), rendering $\gamma$ arbitrary at high $p$.
For VertexCover, inconsistencies at low depth stem from $\beta=0$ being fixed from $p=2$ onward; since the landscape is then $\gamma$-invariant (\cf \autoref{subsec:vertexcover}), $\gamma$ becomes arbitrary and no clear parameter pattern emerges.

When COBYLA is initialised with the sequential parameters (top row), the resulting optimised parameters show similar behaviour, deviating only moderately, but with less smooth changes in $\beta,\gamma$.
Initialisation with $\LRm$ (third row) leads to greater deviations, generally breaking the patterns in both $\beta$ and $\gamma$, except in 10 qubit instances where patterns persist in $\beta$ and in $\gamma$ up to $p\leq3$.
Initialisation with $\LRp$ (second and bottom rows), which corresponds to maximising $H_C$ and thus a disadvantageous starting point, produces deviations larger than with the sequential initialisation.
The largest discrepancies occur in the first and last $\beta,\gamma$ components.
For Max3SAT, optimisation from the $\LRp$ parameters yields slightly greater deviations for smaller instances than in than for $p_\text{target}=21$ MaxCut, though less than when initialised with $\LRm$.
Since these optimised parameters generally yield intermediate average residual energy (\cf~\autoref{subsec:higher_p} and \ref{subsec:max3sat}), and vary little for larger instances, the optimiser is more likely to getting stuck in a local optimum of the \textit{entire} optimisation landscape, that is, the landscape given by all $p$ components of $\beta,\gamma$.
Combined with the observation that large $\beta$ and $\gamma$ values are typically fixed for the final layer (\cf\autoref{subsec:higher_p}, \ref{subsec:max3sat}), this suggests that the optimiser converges to one of the minima located at the edges of the final landscapes.

Notably, none of the methods we examined yields heavily optimised QAOA parameters.
Therefore, we cannot rule out the possibility that there exist better schedules.
Nevertheless, the methods that we examined still often converge towards a state that gives higher quality results than the linear ramp baseline schedules we examined.
This indicates that the parameters given by the more practical approaches we consider in this work, which require lower computational cost compared than heavy optimisation, deviate from smooth patterns.

\section{Discussion}\label{sec:discussion}
Our results paint a nuanced picture of the advantages and disadvantages of QAOA variants: Despite known uniform, generic, and instance-independent characteristics of QAOA at unit depth \(p=1\)~\cite{krueger25,Streif:2020}, 
we find higher circuit depths exhibit an influence of both, the combinatorial \emph{problem}, and the specific \emph{instance}. Nevertheless, commonalities are still observable for deeper circuits, particularly when fixing lower-depth parameters to certain initial values.

Most notably, parameters resulting in low residual energies at every depth lead to a converging landscape that is invariant under $\gamma$ and periodic in $\beta$ with a minimum at $\beta=0$ and maxima towards the top and bottom of the symmetry range.
As \(p\) increases, the choice of \(\gamma\) becomes effectively arbitrary, yielding equivalent energies across all values at large depth.  
Conversely, the optimal choice of $\beta$ approaches zero as $p$ increases.
The number of steps required for the landscapes to converge to such a state depends on the relative magnitudes of $\beta,\gamma$:
For instance, linear ramps with decreasing $\beta$ and increasing $\gamma$ starting from 0 converge slower than methods starting with a non-zero $\gamma \geq \beta$ (\eg the sequential method, or COBYLA using favourable initial parameters).
These \enquote{faster} methods also achieve a better residual energy at lower depth, but may be outperformed by some of the \enquote{slower} converging parameters at higher depth.
This indicates that faster convergence does not necessarily mean better performance.
The sequential method converges most rapidly to a \(\gamma\)-invariant state. Once reached, however, the residual energy remains fixed across subsequent depths, rendering additional QAOA layers ineffective. By contrast, COBYLA initialised with sequential parameters attains a \(\gamma\)-invariant landscape only at \(p_\text{target}\), thereby enabling lower residual energies.  
For linear-ramp parameters, the optimisation landscape does not converge to a \(\gamma\)-invariant form at low depth (\(p_\text{target}=7\) in our experiments), but does so at higher depth (\(p_\text{target}=21\)), thereby achieving lower residual energies than the sequential method. This behaviour indicates that slower convergence at greater depth can yield superior results. Hence, the optimal convergence rate depends on the available circuit depth and should be chosen to ensure that convergence coincides with the final QAOA layer.

The rate of change in the optimisation landscape for a given set of parameters depends not only on the absolute values of \(\gamma\) and \(\beta\), but also on their ratio. We observe that a gradually increasing \(\gamma\) induces a more pronounced vertical shift of the minima towards the region \(\beta=0\) than when \(\gamma\) is initially large and continues to increase.  
With slowly increasing \(\gamma\), the extrema broaden more gradually in the horizontal direction of the parameter landscape when \(\beta\) decreases slowly towards zero. In other words: (i) the rate of propagation under the phase operator governs the speed at which the minima converge towards the centre of the landscape, and (ii) the rate of decrease in \(\beta\) determines the speed at which the landscape becomes horizontally uniform (\ie, \(\gamma\)-invariant).  

Across methods, the sequential approach performs comparably to, and in some cases better than, alternative strategies, depending on problem and circuit depth. Its drawback lies in the diminishing rate of improvement in residual energy as \(p\) increases: At large depth, it is surpassed by methods such as linear ramps that continue to yield improvements throughout the available depths.  

In contrast, linear ramps underperform at shallow depth but continue to improve up to the target depth, with the following exceptions:  
For some problems, notably VertexCover, linear-ramp parameters may fail to yield improvements as \(p\) increases. Moreover, for linear ramps to drive the system towards a low-energy state, the mixer ramp must be aligned with both the initial state and the implementation of \(H_C\). In our setting this requires the initial state \(\ket{+}\) and a problem Hamiltonian \(H_C\) whose minimal eigenvalue encodes the optimum.  
This corresponds to a negative \(\Delta\beta\) (\ie, the \(\LRm\) method), since \(\ket{+}\) is the ground state of \(-U_M\). In contrast, a positive \(\Delta\beta\) (\ie, the \(\LRp\) method) corresponds to \(U_M\), for which \(\ket{+}\) is the \emph{maximum} eigenstate, thereby propagating the system towards a high-energy state.  
COBYLA performance is highly sensitive to the choice of initialisation: parameters that steer the system towards low-energy states provide a favourable starting point, whereas unfavourable choices, such as the \(\LRp\) ramp, are less effective. At shallow depth and for small problem sizes, the optimiser may still identify reasonably good parameters, typically by adjusting mainly the lowest- and highest-depth components of \(\beta\) and \(\gamma\). At greater depth, however, poor initialisation can prevent convergence to low energies. Although COBYLA results often exhibit substantial variance at the target depth, the residual energies across instances are generally consistent~--~except in cases of unfavourable initialisation, where performance can deteriorate.  

For the problems considered in this paper, the best identified parameters frequently deviate from the patterns reported in the literature. This observation is consistent with the convergence behaviour of the optimisation landscape:  
As the residual energy becomes increasingly invariant under \(\gamma\) when \(\beta\) approaches zero, larger deviations from the expected \(\gamma\)-patterns can be tolerated without loss in residual energy. If parameters exist for a given \(p\) that follow the patterns and approximate an eigenstate of \(H_C\) within that depth, which renders the landscape \(\gamma\)-invariant at higher depths, then many other parameter choices may perform comparably well despite substantial deviations from the patterns in their higher-depth components.  
Furthermore, since the adiabatic condition (\ie, propagation is sufficiently slow so as not to disturb adiabatic dynamics) is most restrictive near the minimum spectral gap, parameters may also deviate substantially from the expected patterns in the low- or intermediate-depth components.
Deviations in regions where the spectral gap is large may allow for faster propagation towards the target state without degrading performance.
However, since such parameters still require non-arbitrary values of \(\beta\) and \(\gamma\) to drive the system towards a low-energy state, they are unlikely to deviate as strongly from the patterns as the higher-depth, \(\gamma\)-invariant components. Rather, we expect deviations to arise in the smoothness of the increase or decrease, rather than in the overall continuity of the slope of the parameters.  

Notably, we do not expect the methods examined in this work to produce parameters that strictly conform to the patterns in specific individual instances, as these patterns tend to diminish in the average landscapes shown in \autoref{sec:results}.
This is supported by the relatively low standard deviation and by the fact that the individual landscapes of the examined instances differ only slightly from the corresponding averages across all considered instance-sizes.
For reference, we present the individual landscapes of both the smallest and largest instance of each problem in \autoref{sec:single_instances}.
Furthermore, we also do not attribute deviations form the expected patterns to the possibility that these patterns apply only to larger instances, as the chosen instance sizes are consistet with those used in prior work observing these patterns, particularly Refs.~\cite{zhou_quantum_2020, montanez-barrera_towards_2024}.

\section{Conclusion}\label{sec:conclusion}
We conducted a systematic empirical study of different parametrisation methods in QAOA and their impact on the corresponding optimisation landscapes. Our results demonstrate that parameters, which are optimised in practical settings, can deviate from the prevailing assumption of consistently smooth trends, with \(\beta\) monotonically decreasing and \(\gamma\) monotonically increasing with circuit depth.
We find that such patterns are pronounced only at shallow depth. As \(p\) increases and \(\beta\) decreases, deviations in the higher-depth components of \(\gamma\) become increasingly inconsequential, until in the limit \(\beta=0\) the choice of \(\gamma\) is effectively arbitrary.
Iterative parameter fixing yielded consistent performance across all instances, despite not optimising over \emph{all} components of $\beta,\gamma$ instead of the full \(2p\). While its effectiveness saturates once \(\beta\) converges to zero, the reduced optimisation complexity and strong low-depth performance render it a practical baseline for comparison with more advanced parameter-selection strategies, particularly as only two variational parameters need to be tuned in each iteration.

While we could reproduce some observations of previous work concerning the performance of different parameter selection methods, we could identify circumstances under which these methods may fail:
(1) Linear ramps generally achieve comparable performance across problem instances at high depths, but can be outperformed by alternative methods at shallow depths. Instance-\emph{independent} ramps may lead to substantial reductions in solution quality compared to iteratively fixed or optimised parameters. Moreover, if ramps are misaligned with the initial state or with the implementation of the mixer and phase unitaries, they may effectively approximate the solution of the inverse problem (\eg, maximisation instead of minimisation).  
(2) Optimised QAOA parameters achieve best performance among the methods considered, but incur significantly higher computational cost than fixed parameters or linear ramps. Results exhibit considerable variance and are highly sensitive to the choice of initialisation. Poor initial conditions often hinder matching the performance of ramps or iterative fixing. In this context, iteratively fixed parameters provide a reliable initialisation strategy. 

Our results suggest several promising avenues for improving QAOA. The characteristics outlined in \autoref{sec:discussion} may inform the design of optimiser constraints or motivate variants that iteratively fix parameters with increasing depth. Another direction is to devise fixed linear-ramp methods that adapt to specific problem instances, for example by tuning \(\Delta\beta, \Delta\gamma\) through an optimiser or heuristic rather than assuming complete instance independence. 

\appendices

\section*{Acknowledgement}
We acknowledge support from \programme, grant \grantoth{} (\VE, \MF and \WM). \WM acknowledges support by the \hta.

\section{Landscape Evaluation Algorithm}\label{sec:pseudocode}

\begin{algorithm}
\caption{Landscape scan with iteratively increasing depth}
\label{alg:scan}
\renewcommand{\algorithmicrequire}{\textbf{Input:}}
\renewcommand{\algorithmicensure}{\textbf{Output:}}
\begin{algorithmic}
\REQUIRE{$\mathcal{P}$ = \{problem type, problem size, graph degree, clause to variable ratio, seed\} - parameters for problem specification; \\
$\mathcal{S}$ = \{resolution, $\beta$ bounds, $\gamma$ bounds\} - parameters for scan grid specification; \\
$\mathcal{I}$ = \{$p_\text{start},p_\text{target},p_\text{step}$\} - parameters for iteration loop specification; \\
$\vec{\gamma_\text{init}},\vec{\beta_\text{init}}$ - optional set of parameters which will be fixed instead of the best found values; \\
$opt$ - optimiser with which to optimise $\vec{\gamma_\text{init}},\vec{\beta_\text{init}}$ before starting the scan;}
\ENSURE{$\mathcal{E}$ - List of grids of expectation values corresponding to each $G_p$ evaluated at each iteration}

\STATE $\hat{H_C} \gets \texttt{mapProblemToIsing} (\mathcal{P})$

\IF{$\vec{\gamma_\text{init}},\vec{\beta_\text{init}},opt \neq \text{None}$}
\STATE $res \gets \texttt{QAOA}(\hat{H_C},p_\text{target},\vec{\gamma_\text{init}},\vec{\beta_\text{init}},opt)$
\STATE {\color{white}\(res \gets\)} \texttt{.computeMinimumEigenvalue}()
\STATE $\vec{\gamma_\text{init}},\vec{\beta_\text{init}} \gets res\texttt{.opt\_parameters}$
\ENDIF

\STATE $\vec{\gamma_{b}},\vec{\beta_{b}} \gets \text{None}$
\STATE $\mathcal{E} \gets \texttt{List}()$
\STATE $p \gets p_\text{start}$
\WHILE{$p \leq p_\text{target}$}
\STATE $ansatz \gets \texttt{QAOAAnsatz}(\hat{H_C},p)$
\STATE $G_p \gets \texttt{makeGridofParams} (\mathcal{S},p,\vec{\gamma_{b}},\vec{\beta_{b}})$
\STATE $G_e \gets \texttt{Estimator}(ansatz,G_p,\text{shots}=\text{None})$
\IF{$\vec{\gamma_\text{init}},\vec{\beta_\text{init}} \neq \text{None}$}
\STATE $\vec{\gamma_{b}},\vec{\beta_{b}} \gets \vec{\gamma_\text{init}}[0:p],\vec{\beta_\text{init}}[0:p]$
\ELSE
\STATE $x, y \gets \texttt{getIndex}(\texttt{min}(G_e))$
\STATE $\vec{\gamma_{b}},\vec{\beta_{b}} \gets G_p[x][y]$
\ENDIF
\STATE $\mathcal{E}\texttt{.append}(G_e)$
\STATE $p \gets p + p_\text{step}$
\ENDWHILE
\end{algorithmic}
\end{algorithm}

Algorithm~\ref{alg:scan} lists the procedure for scanning the optimisation landscapes using different parameter selection methods.  
The scan starts by initialising the problem and, in the case that an optimiser has been set, optimises parameters for the target depth $p_\text{target}$, starting from given initial parameters. 
In the case that no optimiser has been set, this step is skipped.
Then, starting from the given start depth, lower-depth parameters are fixed to those of the previous best parameters, the initial parameters for LR, or the optimised parameters if an optimiser is used.
A grid of parameters within the given bounds is then created using Algorithm~\ref{alg:grid}.
Subsequently, a QAOA ansatz from the cost operator and current depth is created.
We then pass the ansatz, cost operator and the grid of parameters to the Qiskit \texttt{Estimator}. The \texttt{Estimator} then binds each of the parameter sets in the grid to a copy of the ansatz and calculates the expectation value of the cost operator in the state prepared by this circuit.
As a result we get a grid of expectation values of the same dimensions as out grid of parameter sets, which we can plot as a landscape.
The best expectation values for each $p$ are used to compute the approximation value plots.

\begin{algorithm}
\caption{makeGridofParams}
\label{alg:grid}
\renewcommand{\algorithmicrequire}{\textbf{Input:}}
\renewcommand{\algorithmicensure}{\textbf{Output:}}
\begin{algorithmic}
\REQUIRE{$n$ - resolution of grid; \\
$B_\text{\{upper, lower\}}(\{\beta, \gamma\})$ - \{upper,lower\} bound on \{$\beta$,$\gamma$\}; \\
$p$ - circuit depth/length of parameter-vectors in the grid; \\
$\vec{\gamma_{f}}$ - the $\gamma$ parameters to be fixed at lower depths; \\
$\vec{\beta_{f}}$ - the $\beta$ parameters to be fixed at lower depths;
}
\ENSURE{$G_p$ - Grid of size $n\times n$ with the points in the grid corresponding to a tuple of parameter-vectors $(\vec{\gamma},\vec{\beta})$ where each parameter-vector has length $p$}

\STATE $G_p \gets \texttt{List}(\text{shape}=(n,n)$
\STATE $intervals_{\beta} \gets B_\text{upper}(\beta) - \frac{i}{n}(B_\text{upper}(\beta) - B_\text{lower}(\beta))\text{ for } i \text{ in } \text{range}(n) $
\STATE $intervals_{\gamma} \gets B_\text{upper}(\gamma) - \frac{i}{n}(B_\text{upper}(\gamma) - B_\text{lower}(\gamma))\text{ for } i \text{ in } \text{range}(n) $
\STATE $i \gets 0$
\WHILE{$i \leq n$}
\STATE $j \gets 0$
\WHILE{$j \leq 0$}
\STATE $\beta_{ij} \gets \vec{\beta_{f}} + intervals_{\beta}[i]$
\STATE $\gamma_{ij} \gets \vec{\gamma_{f}} + intervals_{\gamma}[j]$
\STATE $G_p[i][j] \gets (\beta_{ij},\gamma_{ij})$
\STATE $j \gets j+1$
\ENDWHILE
\STATE $i \gets i+1$
\ENDWHILE

\end{algorithmic}
\end{algorithm}

\section{Max3SAT -- \enquote{Easy} Instances}\label{sec:max3sat_easy}

\begin{figure*}[htb]
  \centering
  \includegraphics{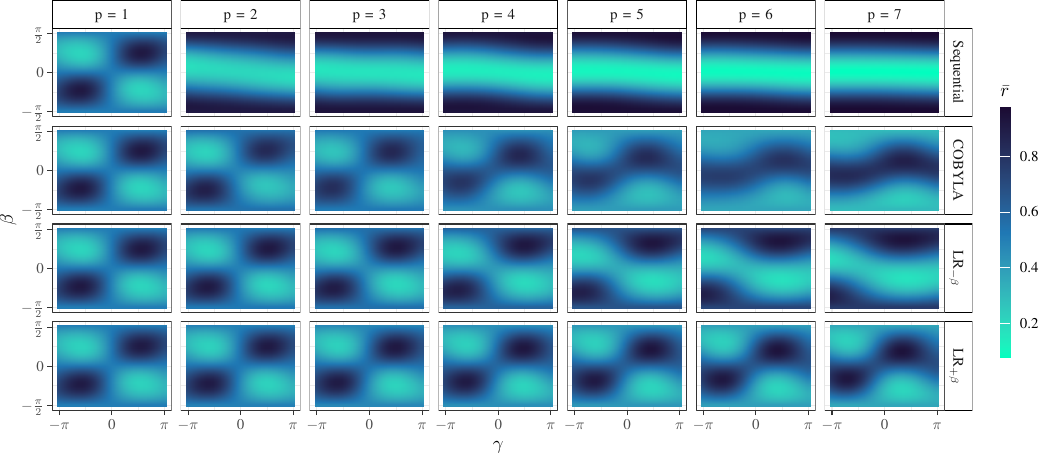}
  \caption{Average residual energy $\bar{r}$ for 10 Max3SAT instances of sizes 14 to 24, with $\alpha\notin(3.5;4.9]$ and $\gamma,\beta$ set to sequentially fixed parameters (top row), optimised parameters using COBYLA starting from the sequential parameters (second row), $\LRm$ parameters with $\Delta_{\beta}=-0.3 , \Delta_{\gamma}=0.6$ (third row), and $\LRp$ parameters with $\Delta_{\beta}=0.3 , \Delta_{\gamma}=0.6$ (bottom row).}
  \label{fig:avg_parameter_scan_max3sat_easy_symmetry}
\end{figure*}

\begin{figure*}[htb]
  \centering
  \includegraphics{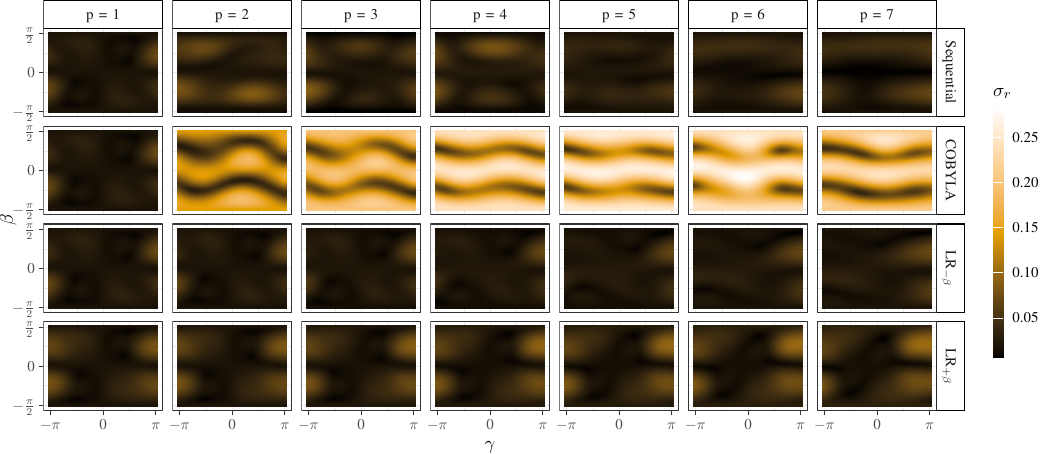}
  \caption{Standard deviation of the residual energy $\bar{r}$ in \autoref{fig:avg_parameter_scan_max3sat_easy_symmetry} over 10 Max3SAT instances.}
  \label{fig:std_parameter_scan_max3sat_easy_symmetry}
\end{figure*}

\begin{figure}[htb]
  \centering
  \includegraphics{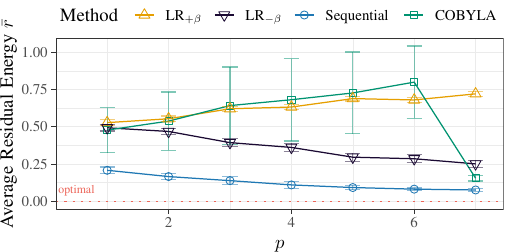}
  \caption{Average and standard deviation of the approximation quality for 10 Max3SAT instances of sizes 14 to 24 with $\alpha\notin(3.5;4.9]$ when fixing parameters at each layer $p$ according to the considered methods.}
  \label{fig:params_parameter_scan_max3sat_easy_symmetry}
\end{figure}

In order to gauge how the landscape changes when we choose a different clause to variable ratio not in the range generally considered as hard ($\alpha\in(3.5;4.9]$), we determined the landscapes resulting from QAOA applied to 10 Max3SAT instances with $\alpha\notin(3.5;4.9]$, which we will refer to as \enquote{easy} instances. The $\bar{r}$ landscape is shown in  \autoref{fig:avg_parameter_scan_max3sat_easy_symmetry} and the $\sigma_r$ landscape is shown in \autoref{fig:std_parameter_scan_max3sat_easy_symmetry}. In both figures, the facets corresponding to the different methods are arranged in the same way as for the hard instance plots (\autoref{fig:avg_parameter_scan_max3sat_hard_symmetry} and \autoref{fig:std_parameter_scan_max3sat_hard_symmetry}).
The average approximation quality of the energies produced by the parameters given by these methods is shown in \autoref{fig:params_parameter_scan_max3sat_easy_symmetry}.

The $\bar{r}$ landscapes produced by the different methods (see \autoref{fig:avg_parameter_scan_max3sat_easy_symmetry}) progress similarly to those of the hard instances, but with a generally lower $\sigma_r$ (see \autoref{fig:std_parameter_scan_max3sat_easy_symmetry}), especially for the landscapes given by the parameters optimised by COBYLA (initialised with $\LRp$).

Notably, all methods perform similarly as in the case, where we chose instances from the (hard) range $\alpha\in(3.5;4.9]$, the main difference being the magnitude of $\sigma_r$.

\section{Typical single instance scans and parameter shape}
\label{sec:single_instances}

In \autoref{sec:results} we focus on the average and standard deviation of the residual energy over multiple instances. 
This appendix supplements the corresponding plots with typical single instances of problems examined in that section, to provide a frame of reference for those results. For the energy landscape scans, we plot the energy expectation value $F_p(\gamma,\beta)$.
The typical instances for each problem are chosen from both, the smallest and largest instances, we examined in the experiments, to show that the landscapes of the individual instances deviate little from the landscapes shown in \ref{sec:results}, and the same convergence patterns can be observed of both the smaller and larger instances.
Taking into consideration the relatively small standard deviation of the energy landscapes discussed in \ref{sec:results}, we conclude that there are likely no outliers that deviate strongly from the observed convergence patterns, but do not show up in the average landscape due to the mean being skewed by other instances.

\subsection{MaxCut}
\label{sec:maxcut_single}

\begin{figure*}[h]
  \centering
	\includegraphics{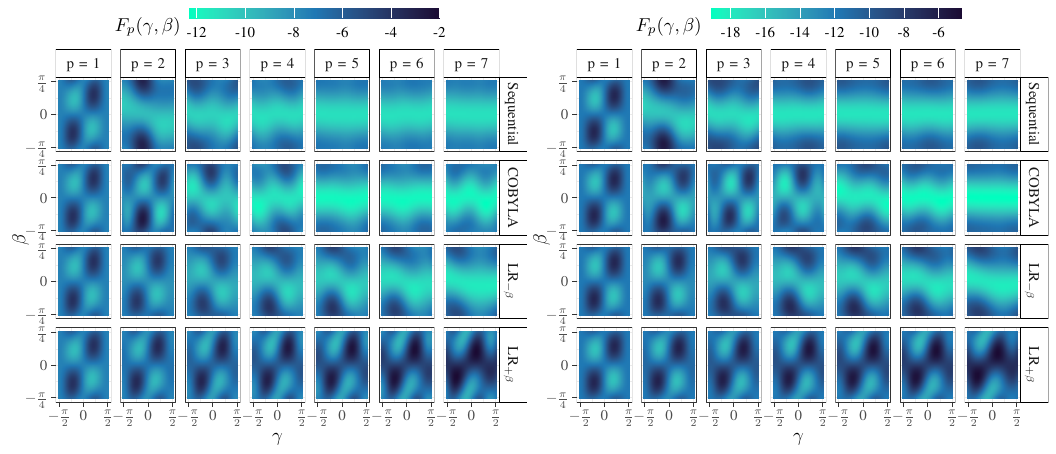}
	\caption{Energy expectation landscapes for one MaxCut instance on a 3-regular 10-vertex  graph (left) and a 3-regular 16-vertex graph (right) with $\gamma,\beta$ set to sequentially fixed parameters (top row), optimised parameters using COBYLA starting from the sequential parameters (second row), $\LRm$ parameters with $\Delta_{\beta}=-0.3 , \Delta_{\gamma}=0.6$ (third row), and $\LRp$ parameters with $\Delta_{\beta}=0.3 , \Delta_{\gamma}=0.6$ (bottom row).}
  \label{fig:maxcut_landscape}
\end{figure*}

\begin{figure*}[htb]
  \centering
	\includegraphics{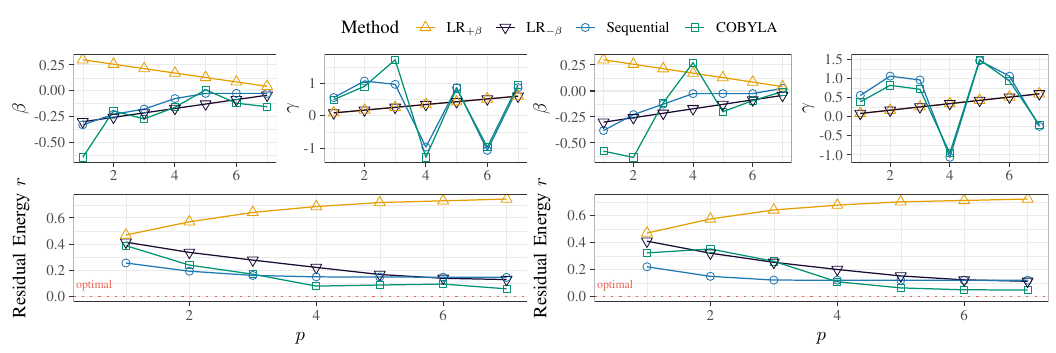}
	\caption{Approximation quality and arrangement of the parameters fixed at each QAOA layer for the 10-qubit (left) and 16-qubit (right) MaxCut instances shown in \autoref{fig:maxcut_landscape}.}
  \label{fig:maxcut_params}
\end{figure*}

\begin{figure*}[htb]
  \centering
	\includegraphics{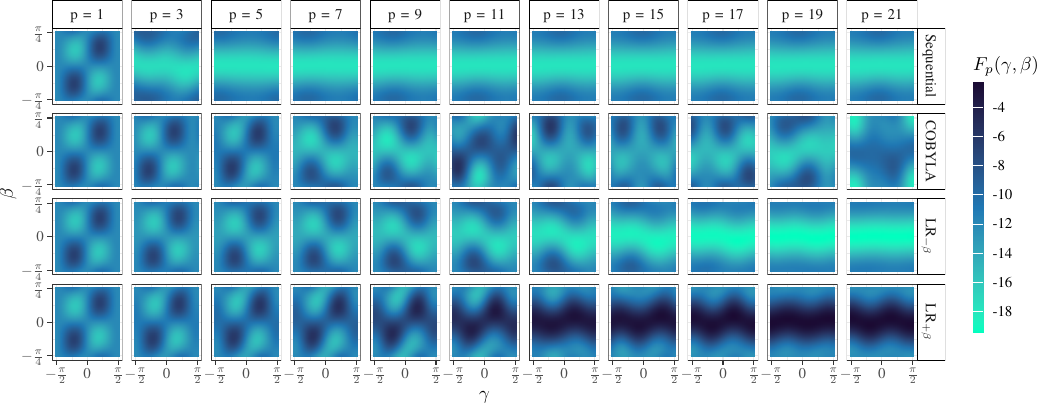}
	\caption{Energy landscapes for the same 3-regular 16-vertex graph instance shown on the right of \autoref{fig:maxcut_landscape} with $\gamma,\beta$ set to sequentially fixed parameters (top row), optimised parameters using COBYLA starting from the $\LRp$ parameters (second row), $\LRm$ parameters with $\Delta_{\beta}=-0.3 , \Delta_{\gamma}=0.6$ (third row), and $\LRp$ parameters with $\Delta_{\beta}=0.3 , \Delta_{\gamma}=0.6$ (bottom row).}
  \label{fig:higher_p_landscape}
\end{figure*}

\begin{figure}[htb]
  \centering
	\includegraphics{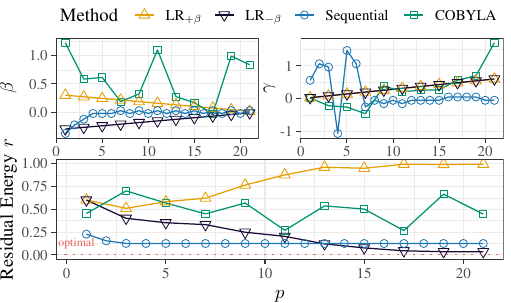}
	\caption{Approximation quality and arrangement of the parameters fixed at each QAOA layer for the MaxCut instance and methods shown in \autoref{fig:higher_p_landscape}. }
  \label{fig:higher_p_params}
\end{figure}

\autoref{fig:maxcut_landscape} and \autoref{fig:maxcut_params} show the results for two of the MaxCut instances examined in \autoref{subsec:maxcut}.
The results for the second MaxCut instance, but at a higher depth (\ie, $p_\text{target}=21$) and incrementing $p$ in steps of 2, are shown in \autoref{fig:higher_p_landscape} and \autoref{fig:higher_p_params}. 

\subsection{VertexCover}\label{sec:vertexcover_single}

\begin{figure*}[h]
  \centering
	\includegraphics{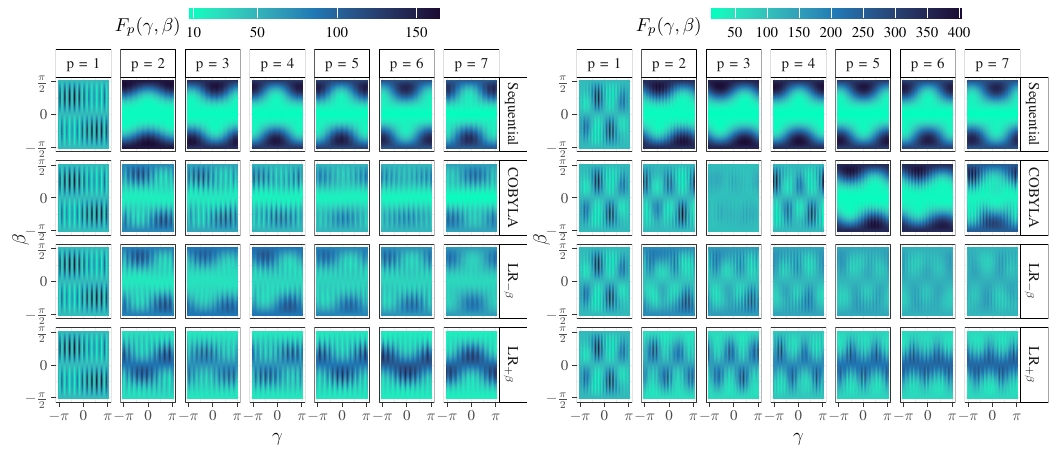}
	\caption{Energy landscapes for one VertexCover instance on a 3-regular 10-vertex  graph (left) and a 3-regular 16-vertex graph (right) with $\gamma,\beta$ set to sequentially fixed parameters (top row), optimised parameters using COBYLA starting from the $\LRm$ parameters (second row), $\LRm$ parameters with $\Delta_{\beta}=-0.3 , \Delta_{\gamma}=0.6$ (third row), and $\LRp$ parameters with $\Delta_{\beta}=0.3 , \Delta_{\gamma}=0.6$ (bottom row).}
  \label{fig:vertcover_landscape}
\end{figure*}

\begin{figure*}[htb]
  \centering
	\includegraphics{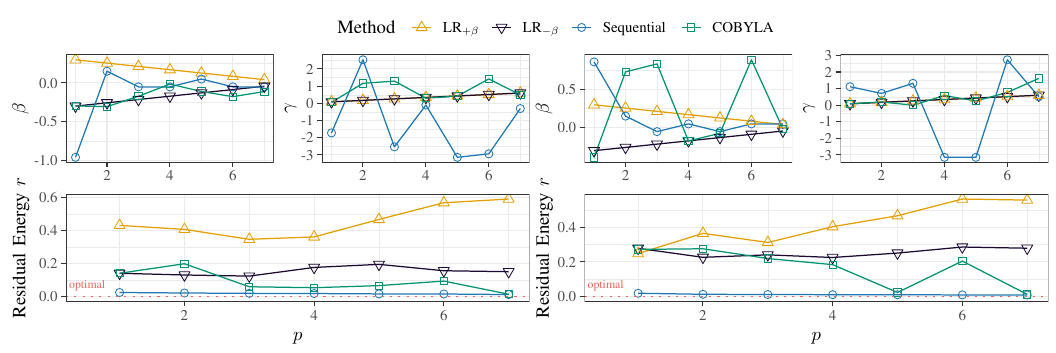}
	\caption{Approximation quality and arrangement of the parameters fixed at each QAOA layer for the 10-qubit (left) and 16-qubit (right) VertexCover instances shown in \autoref{fig:vertcover_landscape}.}
  \label{fig:vertcover_params}
\end{figure*}

\autoref{fig:vertcover_landscape} and \autoref{fig:vertcover_params} correspond to two single instances of the VertexCover problem examined in \autoref{subsec:vertexcover}.
Notably, the density of local minima in the $p=1$ landscape is even more pronounced than in the average landscape shown in \autoref{fig:avg_parameter_scan_vertcover_symmetry}.

\subsection{Max3SAT}\label{sec:max3sat_hard_single}

\begin{figure*}[htb]
  \centering
  \includegraphics{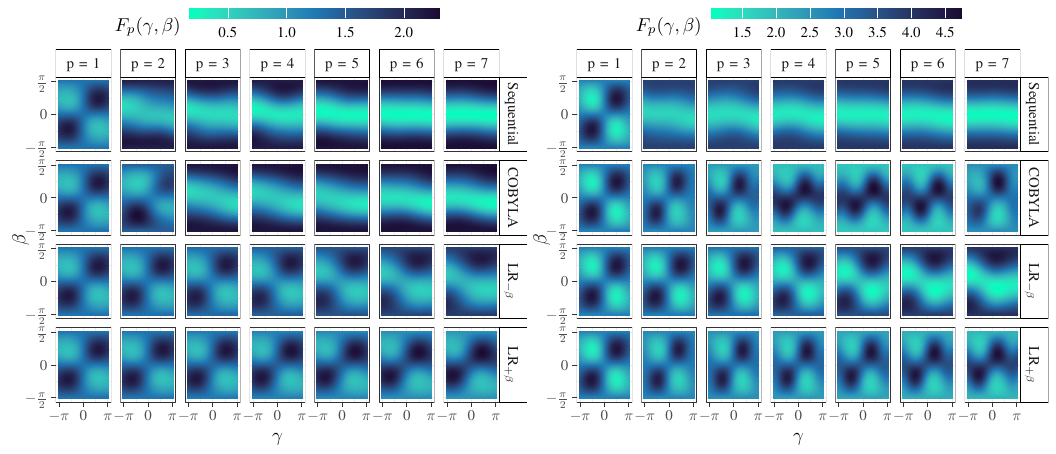}
	\caption{Energy landscapes for one 14-qubit Max3SAT instance with $\alpha=3.\bar{6}\in(3.5;4.9]$ (left) and 24-qubit Max3SAT instance with $\alpha=3.8\in(3.5;4.9]$ (right) with $\gamma,\beta$ set to sequentially fixed parameters (top row), optimised parameters using COBYLA starting from the $\LRp$ parameters (second row), $\LRm$ parameters with $\Delta_{\beta}=-0.3 , \Delta_{\gamma}=0.6$ (third row), and $\LRp$ parameters with $\Delta_{\beta}=0.3 , \Delta_{\gamma}=0.6$ (bottom row).}
  \label{fig:max3sat_landscape}
\end{figure*}

\begin{figure*}[htb]
  \centering
  \includegraphics{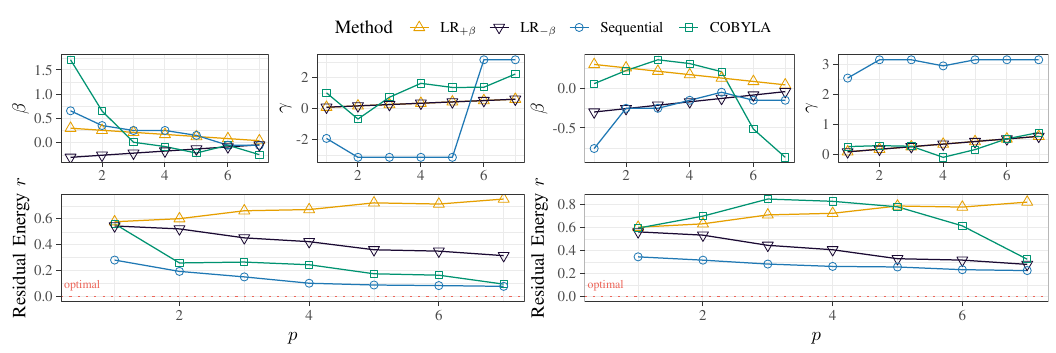}
	\caption{Approximation quality and arrangement of the parameters fixed at each QAOA layer for the 14-qubit (left) and the 24-qubit (right) Max3SAT instances shown in \autoref{fig:max3sat_landscape}.}
  \label{fig:max3sat_params}
\end{figure*}

\autoref{fig:max3sat_landscape} and \autoref{fig:max3sat_params} show the results for two typical Max3SAT instance from the \enquote{hard} range examined in \autoref{subsec:max3sat}. Notably, for the 14-qubit instance, the parameters given by COBYLA seem to converge towards a state where the expectation becomes increasingly invariant under $\gamma$, while the same can not be said for the 24-qubit instance, suggesting that for smaller instances, COBYLA may perform better than the average results shown in \autoref{sec:results}. Aside from that, for the other methods the convergence patterns are similar to what was shown in the average landscapes.

\printbibliography

\EOD{}
\end{document}